\documentclass[epj,nopacs]{svjour}
\usepackage{epsfig,array}
\usepackage{multirow}
\usepackage{graphicx}
\def\nn{\nonumber \\}

\def\fun#1#2{\lower3.6pt\vbox{\baselineskip0pt\lineskip.9pt
\ialign{$\mathsurround=0pt#1\hfil##\hfil$\crcr#2\crcr\sim\crcr}}}

\begin{document}
\newcommand{\be}{\begin{eqnarray}}
\newcommand{\ee}{\end{eqnarray}}
\newcommand{\inli}{\int\limits}
\newcommand{\KL}{\rm\Lambda K^+}
\newcommand{\KS}{\rm\Sigma K}


\date{\today}

\title{\bf  Photoproduction of Baryons Decaying into
N$\pi$ and N$\eta$}

\author{
A.V.~Anisovich \inst{1,2}
\and A.~Sarantsev \inst{1,2}
\and O.~Bartholomy \inst{1}
\and E.~Klempt \inst{1}
\and V.A.~Nikonov  \inst{1,2}
\and U.~Thoma \inst{1,3}
}
\institute{
Helmholtz--Institut f\"ur Strahlen-- und Kernphysik,
Universit\"at Bonn, Germany
\and Nuclear Physics Institute, Gatchina, Russia
\and
Physikalisches Institut, Universit\"at
Gie{\ss}en, Germany
}

\date{Received: \today / Revised version:}


\abstract{A combined analysis of photoproduction data on
$\rm\gamma p\to$ $\rm \pi N$, $\rm\eta N$ was performed including
the data on $\rm K\Lambda$ and $\rm K\Sigma$.
The data are interpreted in an isobar model with $s$--channel
baryon resonances and $\pi$, $\rho$\,($\omega$), $\rm K$, and
$\rm K^*$ exchange in the $t$--channel. Three baryon resonances
have a substantial  coupling to $\rm\eta N$, the well known
$\rm N(1535)S_{11}$, $\rm N(1720)P_{13}$, and $\rm N(2070)D_{15}$.
The inclusion of data with open strangeness reveals the presence
of further new resonances, $\rm N(1840)P_{11}$,
$\rm N(1875)D_{13}$ and $\rm N(2170)D_{13}$.\\
\vspace*{4mm}
}
\vspace{30mm}




\titlerunning{\bf Photoproduction of pions and eta's}

\mail{klempt@hiskp.uni-bonn.de}

\maketitle

\section{Introduction}
The energy levels of bound systems and their decay properties
provide valuable information about the constituents and their
interactions~\cite{Isgur:1995ei}. In quark models, the dynamics of
the three constituent quarks in baryons support a rich spectrum,
much richer than the energy scheme experiments have established so
far~\cite{Capstick:bm,Riska,Metsch}. This open issue is referred
to as the problem of missing resonances. The intense discussion of the 
exotic baryon resonance 
$\rm\Theta^+(1540)$~\cite{Klempt:2004yz,Dzierba:2004db,Hicks:2005gp},
of its existence and of its interpretation, 
has shown limits of the quark model and underlined the need for a deeper
understanding of baryon spectroscopy. Here, the study of
pentaquarks has played a pioneering role, but any new model has to
be tested against the excitation spectrum of the nucleon as well.
The properties of baryon resonances are presently under intense
investigations at several facilities like ELSA (Bonn), GRAAL
(Grenoble), JLab (Newport News), MAMI (Mainz), and SPring-8
(Hyogo). The aim is to identify the resonance spectrum, to
determine spins, parities, and decay branching ratios and thus to
provide constraints for models.
\par
The largest part of our knowledge on baryons stems from pion
induced reactions. In elastic $\rm\pi N$ scattering, the unitarity
condition
provides strong constraints for amplitudes close to the unitarity
limit, since production couplings are related directly to the
widths of resonances and to the cross section. If a resonance has
however a large inelasticity,  its production cross section
in $\rm\pi N$ scattering is
small and it contributes only weakly to the final state. Thus
resonances may conceal themselves from observation in elastic
scattering. This effect could be a reason why the number of
observed states is much smaller than predicted by quark
models~\cite{Capstick:bm,Riska,Metsch}. Information on resonances
coupled weakly to the $\rm \pi N$ channel can be obtained from
photoproduction experiments
and the study of final states different from $\rm \pi N$ such
as multibody final states or final states containing open
strangeness.

The information  from  photoproduction experiments is
complementary to experiments with hadronic beams and gives access
to additional properties like helicity amplitudes. Experiments
with polarised photons provide information which may be very
sensitive to resonances having a small cross section. A clear
example  of such an effect is the observation of the $\rm
N(1520)D_{13}$ resonance in $\eta$ photoproduction. It contributes
very little to the unpolarised cross section but its interference
with $\rm N(1535)S_{11}$ produces a strong effect in the beam
asymmetry. Photoproduction can also provide a very strong
selection tool: combining a circularly polarised photon beam and a
longitudinally polarised target provides a tool to select states
with helicity 1/2 or 3/2 depending on whether the target
polarisation is parallel or antiparallel to the photon helicity.

Baryon resonances have large, overlapping widths rendering
difficult the study of individual states, in particular of those
only weakly excited. This problem can be overcome partly by
looking at specific decay channels. The $\eta$ meson for example
has isospin $I=0$ and consequently, the N$\eta$ final state can
only be reached via formation of N$^*$ resonances. Then even a
small coupling of a resonance to  N$\eta$ identifies it as N$^*$
state. A key point in the identification of new baryon resonances
is the combined analysis of data on photo- (and pion-) induced
reactions with different final states. Resonances must have the
same masses, total widths, and gamma-nucleon couplings,
in all reactions under study. This imposes strong constraints 
for the analysis.

In the present paper we report results of a combined analysis of
photoproduction experiments with $\rm \pi N$, $\rm \eta N$, $\rm
K\Lambda$, and $\rm K\Sigma$  final states. This work is
a first step of a forthcoming analysis of all reactions with
production of baryon resonances in the intermediate state. This
paper concentrates on the reactions $\rm\gamma p\to N\pi$ and $\rm
N\eta$, including available polarisation measurements. Results on
photoproduction of open strangeness are presented in a subsequent
paper~\cite{sarantsev}.

The outline of the paper is as follows: The fit method is
described in section~2, data and fit are compared in section~3. In
section~4 we present the main results of this analysis and discuss
the statistical significance of new 
baryon resonances. Interpretations are offered for the newly found
resonances. The paper ends with a short summary in section 5.

\section{Fit method}
\subsection{Analytical properties of the amplitude and
resonance--Reggeon duality}

The choice of amplitudes used to describe the data is partly
driven by experimental observations. In pion photoproduction,
angular distributions exhibit strong variations indicating the
presence of baryon resonances. On the other hand, all data on
single--meson photoproduction have pro\-minent forward peaks in the
region above 2000 MeV which can be associated with $t$--channel
exchange processes. Reg\-ge behaviour, extrapolated to the
low--energy region, describes the cross section in the resonance
region ``on average". This feature is known as Reggeon--resonance
duality (see \cite{Duality1} and references therein). It gave hope
for a self--consistent construction of hadron--hadron interactions
in both, the low--energy and the high--energy region. However there
is a problem with unitarity: The $s$--channel unitarity
corrections destroy the one--Reggeon exchange picture, while the
$s$--channel resonance amplitudes do not satisfy the $t$--,
$u$--channel unitarity \cite{Duality2}. So it seems reasonable to
extract the resonance  structure of the amplitude together with
phenomenological reggeized $t$-- and $u$--channel exchange
amplitudes.

The scattering amplitude has the following analytical properties.
The partial--wave or multipole amplitudes contain singularities
when the scattering particles can form a bound state with mass
$M$. Unstable bound states with a finite width $\Gamma$ have a pole
singularity at $s= M^2 - i\Gamma M$ in the complex plane. At the
opening of  thresholds, the amplitude  acquires a square root
singularity (right--hand singularity); $t$--exchange leads to
left--hand singularities at $t = \mu^2$ (one--particle exchange with
mass $\mu$), $t = 4\mu^2$ (exchange of two of these
particles) and so on. In three--body interactions the three--particle
rescattering amplitude gives a triangle singularity which may
contribute significantly to the cross section under some particular
kinematical conditions \cite{triangle}. Triangle singularities grow
logarithmically and are thus weaker than a pole or a threshold
singularity. In most cases, triangle singularities can be accounted
for by introducing phases to resonance couplings.

In our present analysis, the primary goal is to get information
about the leading (pole) singularities of the photoproduction
amplitude. For this purpose, a representation of the amplitude as
a sum of $s$--channel resonances together with some $t$-- and
$u$--exchange diagrams is an appropriate representation. Strongly
overlapping resonances are parameterised as $K$--matrix. In many
cases it is sufficient to use a relativistic Breit--Wigner
parameterisation, though.

We emphasise that the amplitudes given below satisfy gauge
invariance, analyticity and unitarity. However, when $t$--, $u$--,
and $s$--channel amplitudes are added, unitarity is violated. In
principle, this can be avoided by projecting the $t$-- and
$u$--channel amplitudes onto $s$--channel amplitudes of defined
spins and parities. The projected amplitudes are however small, and
the violation of unitarity is mild as long as $t$-- and
$u$--channel amplitudes contribute only a small fraction to the
total cross section. In this analysis, amplitudes for
photoproduction of baryon resonances and their decays are
calculated in the framework of relativistic tensor operators. The
formalism is fully described in~\cite{Anisovich:2004zz}; here
parameterisations of resonances used under different conditions
are given.

\subsection{Parameterisations of resonances}

The differential cross section for production of two or more
particles has the form:
\be 
{\rm d\sigma}=\frac{(2\pi)^4
|A|^2}{4\,\sqrt{(k_1k_2)^2-m_1^2m_2^2}}\,
d\Phi_n(k_1+k_2,q_1,\ldots,q_n)
\ee
where $k_i$ and $m_i$ are the
four--momenta and masses of the initial particles (nucleon and
$\gamma$ in the case of photoproduction) and $q_i$ are the
four--momenta of final state particles. 
$d\Phi_n(k_1+k_2,q_1,\ldots,q_n)$ is the n--body phase volume
\be
&&d\Phi_n(k_1+k_2,q_1,\ldots,q_n) = \nn && \delta^4(k_1+k_2-
\sum\limits_{i=1}^n q_i) \prod\limits_{i=1}^n
\frac{d^3q_i}{(2\pi)^3 2q_{0i}}
\ee
where $q_{0i}$ is time component (energy).
The differential cross section
for photoproduction of single mesons is given by 
\be
{\rm d\sigma}=\frac{\sqrt{(s-(m_\mu+m_B)^2))(s-(m_\mu-m_B)^2)}} {16\pi
s(s-m_N^2)}|A|^2 \nn 
\ee 
where $s= (k_1 + k_2)^2 = (q_1 + q_2)^2$
is the square of the total energy, $m_\mu,$ ($\mu=\rm
\pi,\eta,K$), $m_B$ ($B=\rm N,\Lambda,\Sigma$) the meson and
baryon masses, respectively.

The $\eta$ photoproduction cross section is dominated by  $\rm
N(1535)S_{11}$. It overlaps with $\rm N(1650)S_{11}$ and the two
$\rm S_{11}$ resonances are described as two--pole, four--channel
$K$--matrix ($\rm \pi N$, $\rm \eta N$, $\rm K\Lambda$ and $\rm
K\Sigma$). The photoproduction amplitude can be written in the
$P$--vector approach since the $\gamma\rm N$ couplings are weak and
do not contribute to rescattering. The amplitude is then given by
\be A_a \;=\; \hat P_b\;(\hat I\;-\;i\hat \rho \hat
K)^{-1}_{ba}\,. \ee The phase space $\hat \rho$ is a diagonal
matrix with \be \rho_{ab}\;=\;\delta_{ab}\;\rho_a,\;\;
a,b=\pi\mbox{$\rm N, \eta N, K \Lambda, K \Sigma$}. \ee and \be
\rho_{a}(s)=\frac{\sqrt{(s-(m_\mu+m_B)^2))(s-(m_\mu-m_B)^2)}}{s}.
\label{phv} \ee The production  vector $\hat P$ and the $K$--matrix
$\hat K$ have the following parameterisation:
 \be
 K_{ab}\;=\;\sum_\alpha \frac{g_a^{(\alpha)} g_b^{(\alpha)}}
{M^2_\alpha - s} \;+\; f_{ab},
 \;\;\;
 P_{b}\;=\;\sum_\alpha \frac{
g_{\gamma \rm N}^{(\alpha)} g_b^{(\alpha)}}{M^2_\alpha - s} \;+\;
 \tilde f_{b} \nn
 \ee
 where $M_\alpha$, $g_a^{(\alpha)}$ and $g_{\gamma\rm N}^{(\alpha)}$
  are the mass, coupling constant and production constant of the resonance
 $\alpha$; $f_{ab}$ and $\tilde f_{b}$ are non--resonant terms.

Other resonances were taken as Breit--Wigner amplitude:
\be
A_a=\frac{g_{\gamma N} \tilde g_{a}(s)}
{M^2-s-i\;M\tilde \Gamma_{tot}(s)}~~~~~~~~~~~~
\ee
States with masses above 1700\,MeV were parameterised with a constant
width to fit exactly the pole position. For resonances below 1700\,MeV,
$\tilde\Gamma_{tot}(s)$ was parameterised by
\be
\tilde\Gamma_{tot}(s)=\Gamma_{tot} \frac{\rho_{\pi N}(s) k^{2L}_{\pi
N}(s) F^2(L,k^2_{\pi N}(M^2),r)}
     {\rho_{\pi N}(M^2) k^{2L}_{\pi N}(M^2) F^2(L,k^2_{\pi N}(s),r)}\;,
\nonumber \\
k^2_{a}(s)=\frac{(s-(m_\mu+m_B)^2))(s-(m_\mu-m_B)^2)}{4s}\;.
\ee
Here,  $L$ is the orbital momentum and $k$ is the relative momentum
for the decay into $\pi N$ ($\mu=\pi$, $B=N$). 
$F(L,k^2,r)$ are Blatt--Weiskopf form factors, taken with a radius
$r=0.8$\,fm. The exact form of these factors can be found e.g. in
\cite{Anisovich:2004zz}. $g_{\gamma N}$ is the production coupling
and $\tilde g_{a}$ are decay couplings of the resonance into meson
nucleon channels. These couplings are suppressed at large energies by
a factor
\be
\tilde g_{a}(s) =g_{a}
\sqrt{\frac{\rm 1.5\,GeV^2}{{\rm 1.0\,GeV^2}+k_{a}^2}}\,.
\ee
The factor proved to be useful for two--meson photoproduction. For
photoproduction of single mesons, it plays almost no role and is
only introduced here for the sake of consistency.

The partial widths
are related to the couplings as
\begin{eqnarray}
M \Gamma_{a} &=&\tilde g_{a}^2
\frac{\rho_{a}(M^2) k_{M^2}^{2L}}{F^2(L,k_{M^2}^2,r)}
\frac{m_B+\sqrt{m_B^2+k_{a}^2}}{2m_B}\beta_L\;,
\nonumber\\
\beta_L&=&\frac 1L\prod\limits_{l=1}^{L}\frac{2l-1}{l} ,\;
\quad\quad\quad J=L-\frac 12\;,
\nonumber\\
\beta_L&=&\frac 1{2L+1}\prod\limits_{l=1}^{L}\frac{2l-1}{l} ,\quad
J=L+\frac 12\;.
\end{eqnarray}
Here $J$ is the total momentum of the state.

\subsection{$\bf t$-- and $\bf u$--channel exchange parameterisations}

At high energies, there are clear peaks in the forward direction of
photoproduced mesons.
The forward peaks are connected with meson exchanges in the
$t$--channel. These contributions are parameterised as
$\pi$, $\rho(\omega)$, $\rm K$, and $\rm K^*$ exchanges.

These contributions are reggeized by using \cite{collins}
\be
T(s,t)=g_1(t)g_2(t)\frac{1+\xi exp(-i\pi\alpha(t))}{\sin(\pi\alpha(t))}
\left (\frac{\nu}{\nu_0} \right )^{\alpha(t)}\,,
\nonumber\\
\nu=\frac 12 (s-u).
\phantom{12444444444444444444}
\ee
Here, $g_i$ are vertex functions, $\alpha(t)$ is a function
describing the trajectory, $\nu_0$ is a normalisation factor (which can
be taken to be 1) and $\xi$ is the signature of the trajectory.
Exchanges of
$\pi$ and $\rm K$ have positive, $\rho$, $\omega$, and $\rm
K^*$ exchanges have negative signature.

\begin{table}[b!]
\caption{Data used in the partial wave analysis,
$\chi^2$ contributions and fitting weights.}
\renewcommand{\arraystretch}{1.3}
\begin{center}
\begin{tabular}{lccccr}
\hline\hline
 Observable & $N_{\rm data}$   & $\chi^2$ & $\chi^2/N$   & Weight&
 Ref. \\
\hline \hline
$\rm\sigma(\gamma p \rightarrow p\pi^0)$
 & 1106 &  1654  &  1.50 &  8    &
\cite{Bartholomy:04} \\
$\rm\sigma(\gamma p \rightarrow p\pi^0)$
 &  861 &  2354  &  2.74 &  3.5  &
 \cite{GRAAL1} \\
$\Sigma(\rm\gamma p \rightarrow p\pi^0)$
 &  469 &  1606  &  3.43 &  2    &
 \cite{GRAAL1}\\
$\Sigma(\rm\gamma p \rightarrow p\pi^0)$
 &  593 &  1702  &  2.87 &  2    &
 \cite{SAID1}\\
$\rm\sigma(\gamma p \rightarrow n\pi^+)$
 & 1583 &  4524  &  2.86 &  1    &
\cite{SAID2} \\
$\rm\sigma(\gamma p \rightarrow p\eta)$
 &  100 &   158  &  1.60 &  7    &
 \cite{Krusche:nv}\\
\hline
$\rm\sigma(\gamma p \rightarrow p\eta)$
 &  667 &   608  &  0.91 & 35    &
\cite{Crede:04} \\
$\Sigma(\rm\gamma p \rightarrow p\eta)$
 &   51 &   114  &  2.27 & 10    &
 \cite{GRAAL2}\\
$\Sigma(\rm\gamma p \rightarrow p\eta)$
 &  100 &   174  &  1.75 & 10    &
 \cite{GRAAL1}\\
\hline
$\rm\sigma(\gamma p \rightarrow \Lambda K^+)$
 &  720 &   804  &  1.12 &  4    &
\cite{Glander:2003jw} \\
$\rm\sigma(\gamma p \rightarrow \Lambda K^+)$
 &  770 &  1282  &  1.67 &  2    &
\cite{McNabb:2003nf} \\
$\rm P(\gamma p \rightarrow \Lambda K^+)$
 &  202 &   374  &  1.85 &  1    &
\cite{McNabb:2003nf} \\
$\Sigma(\rm\gamma p \rightarrow \Lambda K^+)$
 &   45 &    62  &  1.42 & 15    &
\cite{Zegers:2003ux} \\
\hline
$\rm\sigma(\gamma p \rightarrow \Sigma^0 K^+)$
 &  660 &   834  &  1.27 &  1    &
\cite{Glander:2003jw} \\
$\rm\sigma(\gamma p \rightarrow \Sigma^0 K^+)$
 &  782 &  2446  &  3.13 &  1    &
\cite{McNabb:2003nf} \\
$\rm P(\gamma p \rightarrow \Sigma^0 K^+)$
 &   95 &   166  &  1.76 &  1    &
\cite{McNabb:2003nf} \\
$\Sigma(\rm\gamma p \rightarrow \Sigma^0 K^+)$
 &   45 &    20  &  0.46 & 35    &
\cite{Zegers:2003ux} \\
\hline
$\rm\sigma(\gamma p \rightarrow \Sigma^+ K^0)$
 &   48 &   104  &  2.20 &  2    &
\cite{McNabb:2003nf} \\
$\rm\sigma(\gamma p \rightarrow \Sigma^+ K^0)$
 &  120 &   109  &  0.91 &  5    &
\cite{Lawall:2005np} \\
\hline
\hline
\end{tabular}
\end{center}
\label{chi}
\renewcommand{\arraystretch}{1.0}
\end{table}
For $\rho$($\omega$) exchange, $\alpha(t)=0.50+0.85t$. The pion
trajectory is given by $\alpha(t)=-0.014+0.72t$, the $\rm K^*$ and
$\rm K$ trajectories are represented by
$\alpha(t)=0.32+0.85t$ and $\alpha(t)=-0.25+0.85t$, respectively.
The full expression for the $t$--channel amplitudes can be found
in \cite{Anisovich:2004zz}.

The $u$--channel exchanges were parameterised as nucleon,
$\rm\Lambda$, or $\rm\Sigma$ exchanges.


\begin{figure}[b!]  
\vspace*{-6mm}
\centerline{\hspace*{-6mm}
\includegraphics[width=0.49\textwidth]{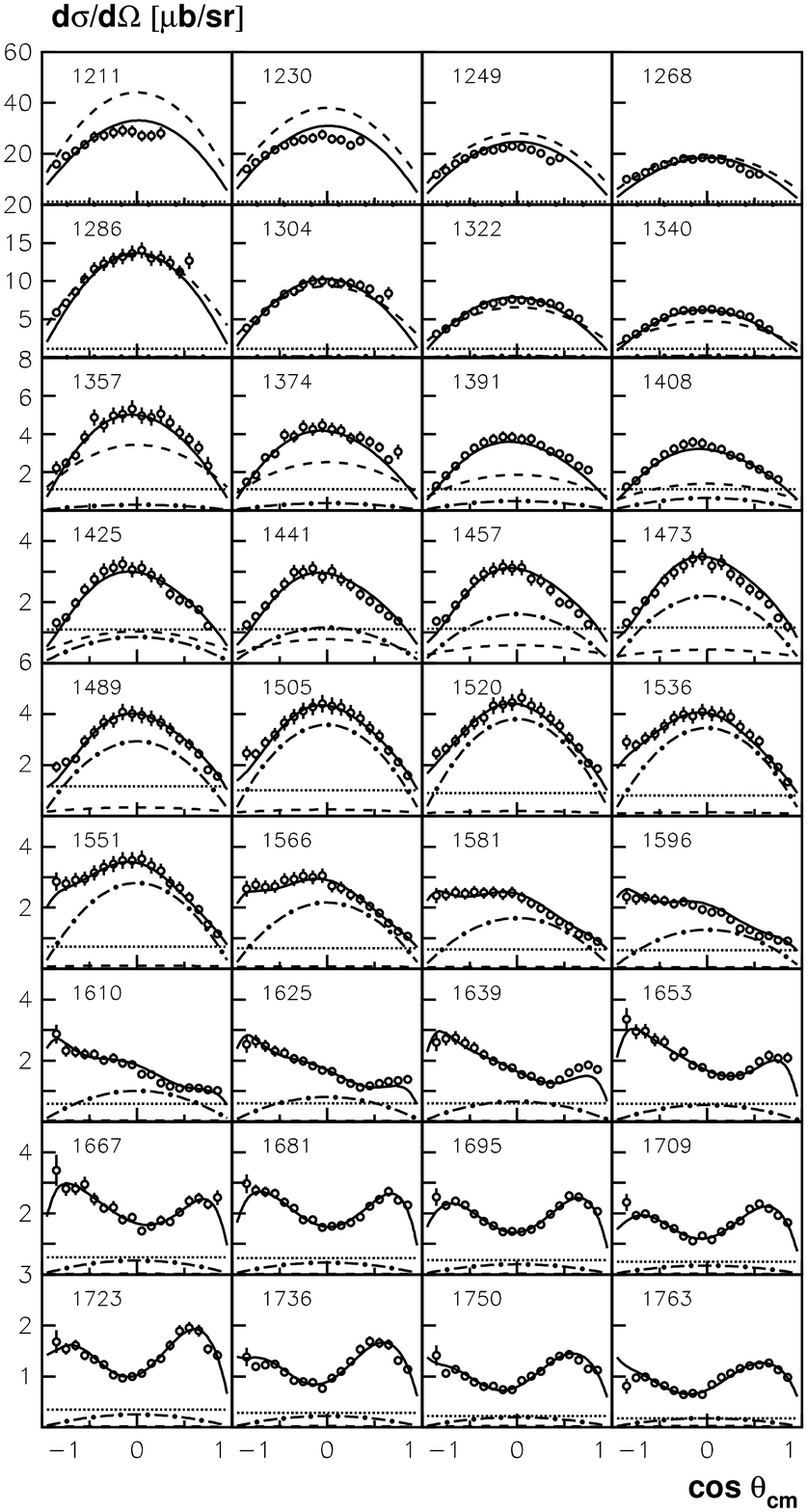}}
\vspace*{8mm}
\end{figure}
\begin{figure}[b!]  
\vspace*{-6mm}
\centerline{\hspace*{-4.2mm}
\includegraphics[width=0.49\textwidth]{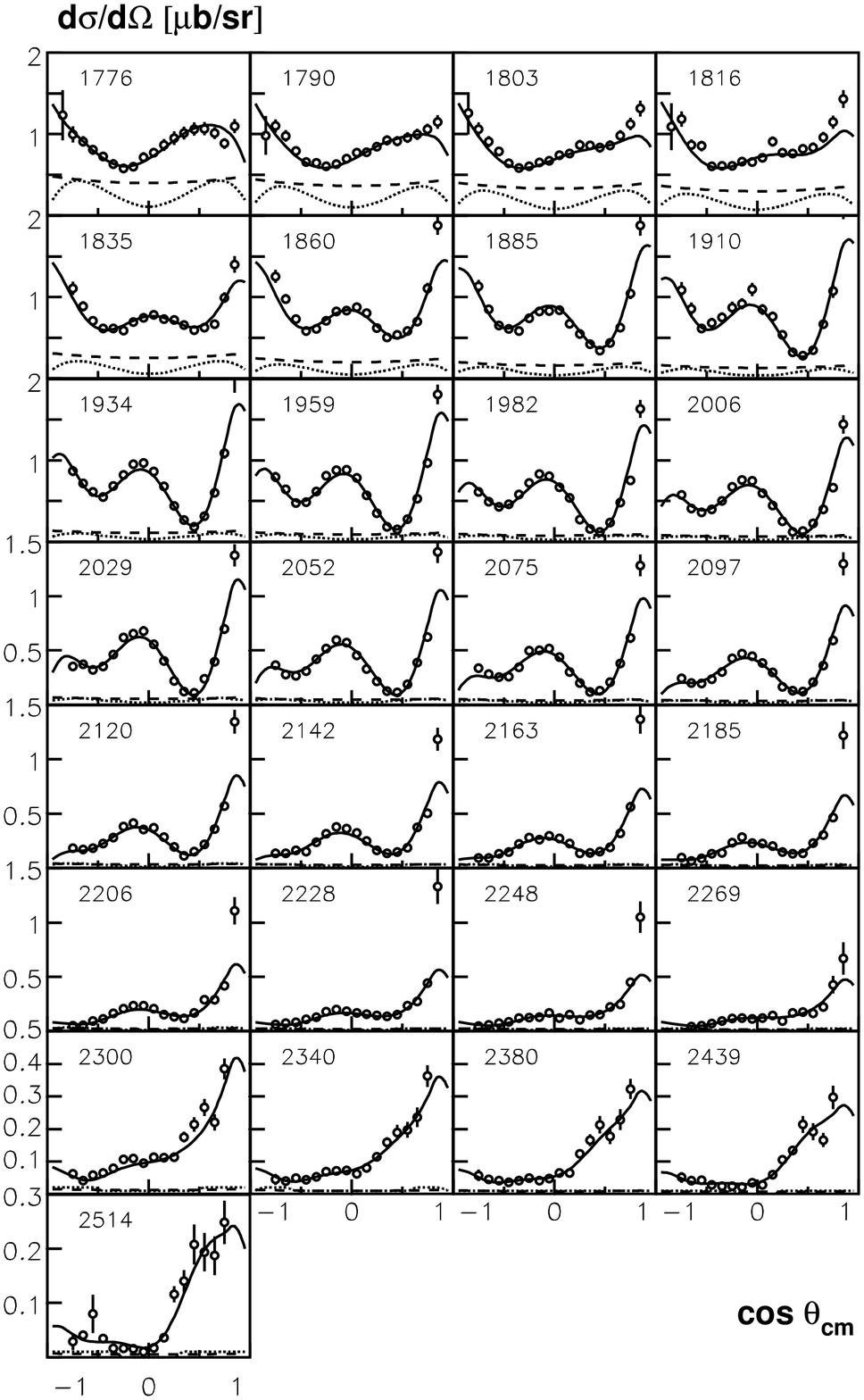}}
\vspace*{-20mm}
\caption{\label{cbelas_pi0}
Differential cross section for $\rm \gamma p \rightarrow p\pi^0$ from
CB--ELSA and PWA result (solid line). The left part of the figure
shows the contribution of $\rm \Delta(1232)P_{33}$ together with
non--resonant background (dashed line), the two $\rm S_{11}$
resonances (dotted line) and $\rm N(1520)D_{13}$ (dash-dotted
line); in the right figure, the contributions of $\rm
\Delta(1700)D_{33}$ (dashed line) and $\rm N(1680)F_{15}$
(dotted line) are shown.
}
\end{figure}

\section{Fits to the data}
In this paper, we report results on baryon resonances and their
coupling to $\rm N\pi$ and  $\rm N\eta$. The results are based on a
coupled--channel analysis of various data sets on photoproduction
of different final states. The data comprise CB--ELSA $\pi^0$ and
$\eta$ photoproduction data~\cite{Bartholomy:04,Crede:04}, the
Mainz--TAPS data~\cite{Krusche:nv} on $\eta$ photoproduction,
beam--asym\-metry measurements of $\pi^0$ and
$\eta$~\cite{GRAAL1,SAID1,GRAAL2},  and data on $\rm\gamma
p\rightarrow n\pi^+$~\cite{SAID2}. The high precision  data from
GRAAL~\cite{GRAAL1} do not cover the low mass region; therefore we
extract further data from the compilation of the  SAID database
\cite{SAID1}. This data allows us to define the ratio of helicity
amplitudes for the $\rm\Delta(1232)P_{33}$ resonance.

Data on photoproduction of $\rm K^+\Lambda$, $\rm K^+\Sigma$,
and  $\rm K^0\Sigma^+$ from SAPHIR~\cite{Glander:2003jw} and
CLAS~\cite{McNabb:2003nf},
and beam asymmetry data for $\rm K^+\Lambda$, $\rm K^+\Sigma$
from LEPS~\cite{Zegers:2003ux} are also included in the coupled--channel
analysis. The results on couplings of baryon resonances to $\rm K^+\Lambda$
and $\rm K^+\Sigma$ are documented in a separate
paper~\cite{sarantsev}.

The fit uses 14\, N$^*$ resonances coupling to  N$\pi$, N$\eta$,
$\rm K\Lambda$, and  $\rm K\Sigma$ and 7 $\rm \Delta$ resonances
coupling to N$\pi$ and $\rm K\Sigma$. Most resonances are
described by relativistic Breit--Wigner amplitudes. For the two
S$_{11}$ resonances at 1535 and 1650\,MeV, a four--channel
$K$--matrix ($\rm N\pi$, $\rm N\eta$, $\rm K\Lambda$, $\rm K\Sigma$) 
is used. The
background is described by reggeized $t$--channel $\pi$,
$\rho$\,($\omega$), $\rm K$ and $\rm K^*$ exchanges and by baryon
exchanges in the $s$-- and $u$--channels.

\begin{figure}[t!] 
\begin{center}
\vspace{-2mm}
\includegraphics[width=0.47\textwidth,height=0.60\textheight]{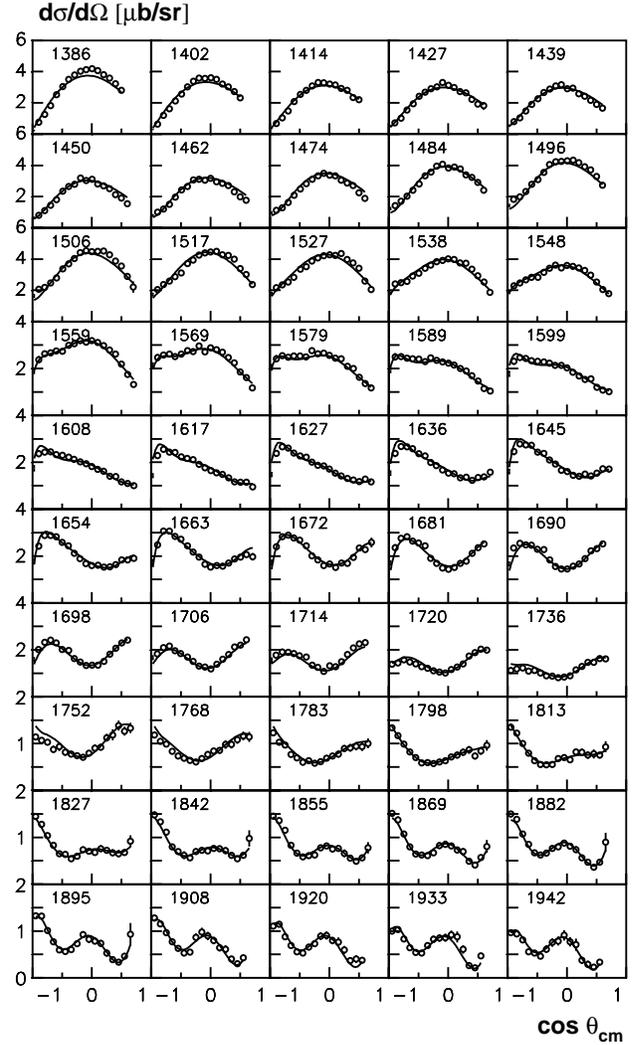}
\vspace*{-0mm}
\caption{\label{graal04_pi0}
Differential cross section for $\rm
\gamma p \rightarrow p\pi^0$ from GRAAL and PWA result (solid
line).}
\vspace*{-7mm}
\end{center}
\end{figure}
The $\chi^2$ values for the final solution of the partial--wave
analysis are given in Table~\ref{chi}. Weights are given to the
different data sets
included in this analysis with which they enter the fits.
In the choice of
weights, some judgement is needed. The CB-ELSA data on pion and
$\eta$ photoproduction are the main source of the analysis and
thus have large weights. The beam polarisation measurements for
open strangeness production are also emphasized as discussed
in  \cite{sarantsev}.
Fits were performed with a variety of
different weights; accepted solutions resulted not only in a good
overall $\chi^2$; emphasis was laid on having a good fit of all
data sets. Changing the weights may result in pictures showing
larger discrepancies; the changes of pole positions are only
small.
\par
The fit minimises a pseudo--chisquare function which we call
$\chi^2_{\rm tot}$. It is given by
\be
\chi^2_{\rm tot}=\frac{\sum w_i\chi^2_i}{\sum w_i\,N_i}\,\sum N_i
\label{chi_tot}
\ee
where the $N_i$ are given as $N_{\rm data}$ (per channel) in the
$2^{\rm nd}$ column of Table~\ref{chi} and the weights in the last
column.

\subsection{Fit to the \boldmath $p\rm\pi^0$ \unboldmath data}

The differential cross sections for the CB--ELSA
$\gamma p \rightarrow p\pi^0$ data are shown in Fig.~\ref{cbelas_pi0}.
The main fit is
represented as solid line. The figure also shows the most
important individual contributions. The contribution of
$\rm\Delta(1232)$ (given as dashed line, on the left panel) dominates
the low--energy region, for small photon energies it even exceeds
the experimental cross section, thus underlining the importance of
interference effects. Non--resonant background amplitudes, given
by a pole at $s\sim -1$\,GeV$^2$ and by a $u$--channel exchange
diagram, are needed to describe the shape of the $\rm\Delta(1232)$.
The pole at negative $s$ represents the left--hand cuts.
\par
The two $\rm S_{11}$ resonances at 1535 and at 1650\,MeV are described
as $K$--matrix. Their sum is depicted as dotted line. The $\rm S_{11}$
contribution is flat in $\rm\cos\Theta_{\rm cm}$. The contribution of
the $\rm D_{13}(1520)$ shown as dash--dotted line in
Fig.~\ref{cbelas_pi0} (left panel). It is strong in the
$1400-1600$\,MeV mass region. At higher energies
(Fig.~\ref{cbelas_pi0}, right panel) the most significant
contributions come from $\rm\rm\Delta(1700)D_{33}$ (dashed line) and
from $\rm N(1680)F_{15}$ (dotted line).
For invariant $\rm p\gamma$ masses above 1800\,MeV, the most forward point
in Fig.~\ref{cbelas_pi0} is not reproduced by the fit. If this point
is given a very small error (to ensure that the fit describes these
points), the overall agreement between data and fit becomes somewhat
worse; resonance masses and widths change by a few MeV, at most.
\par
\begin{figure}[t!] 
\centerline{\includegraphics[width=0.50\textwidth]{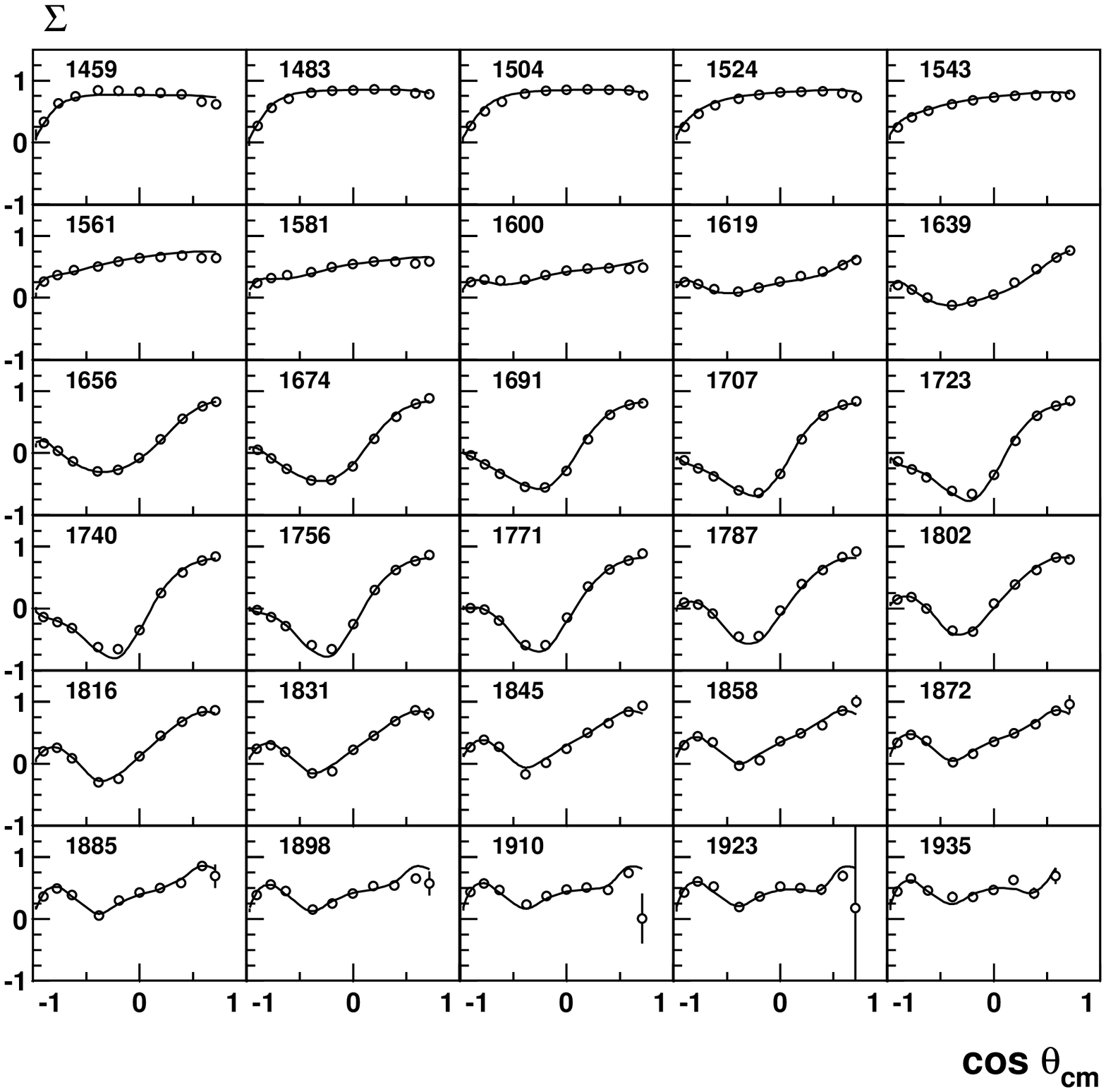}}
\vspace*{-5mm}
\caption{Photon beam asymmetry $\Sigma$ for
$\rm\gamma p \rightarrow p\pi^0$ from GRAAL \cite{GRAAL1} and PWA
result (solid line). }
\label{graal04_pi0_sigma}
\vspace*{4mm}
\centerline{\includegraphics[width=0.50\textwidth,height=0.54\textheight,clip]{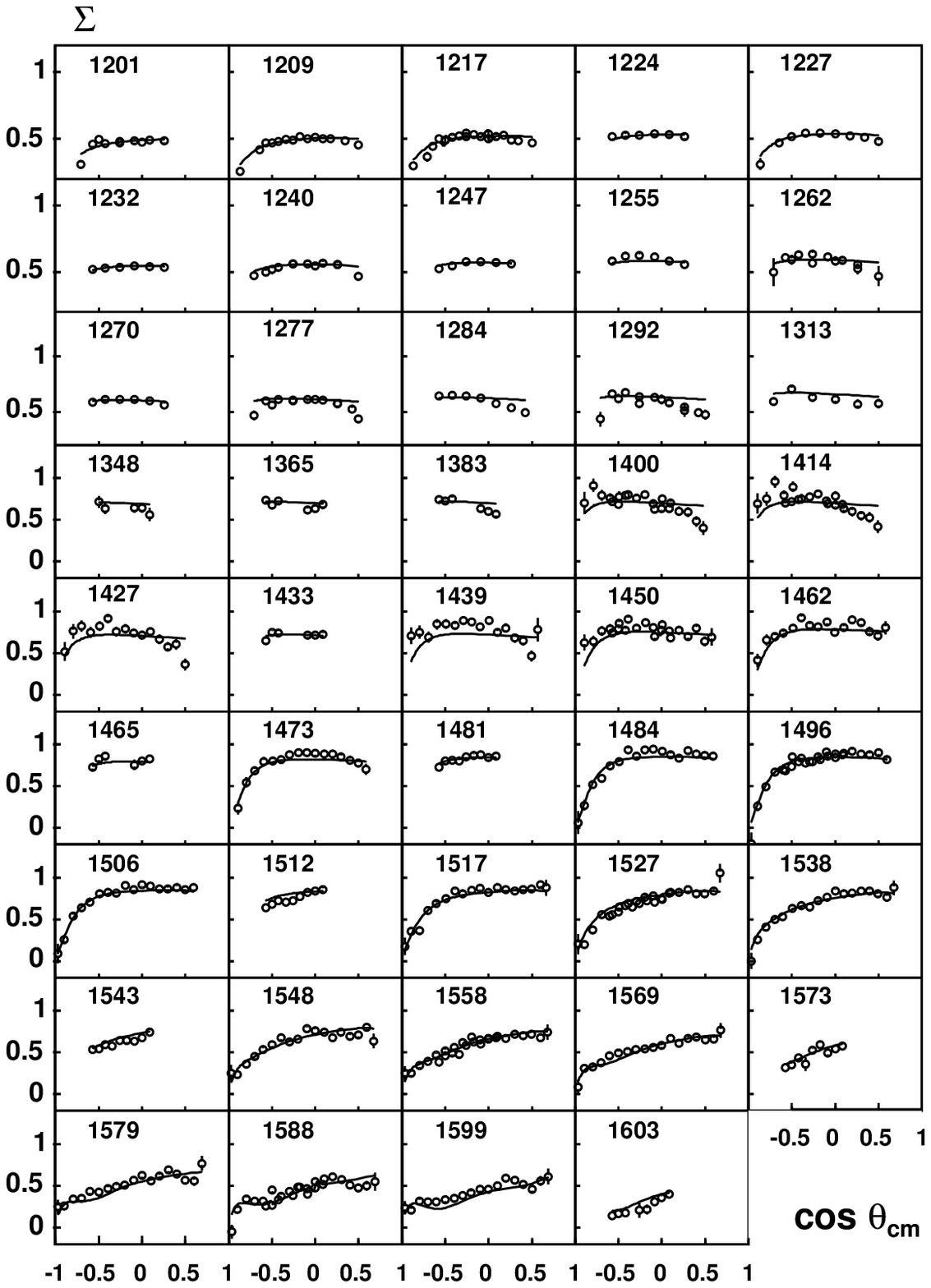}}
\vspace*{-2mm}
\caption{\label{sigma_said}
Photon beam asymmetry $\Sigma$ for
$\rm \gamma p \rightarrow p\pi^0$ from \cite{SAID1}
and PWA result (solid line).}
\vspace*{-6mm}
\end{figure}

Recent data from GRAAL \cite{GRAAL1} on the differential cross section
for  $\rm\gamma p \rightarrow p\pi^0$ and on the photon beam asymmetry
$\Sigma$ are compared to our fit in Figs.~\ref{graal04_pi0}
and \ref{graal04_pi0_sigma}; older  beam asymmetry data are 
shown in Fig.~\ref{sigma_said}.

\subsection{Fit to \boldmath $\rm n \pi^+$ \unboldmath
photoproduction data}
\vspace*{-1mm}

It is important to include data on  $\rm n\pi^+$ photoproduction
since the combination of the  $\rm n\pi^+$  and  $\rm p\pi^o$
channels defines the isospin of $s$-channel baryons. Without this
information, pairs of resonances like $\rm N(1700)D_{13}$ and $\rm
\Delta(1700)D_{33}$ cannot be separated. A fit with both having
large destructively interfering amplitudes may give a good
$\chi^2$ even though the fit is physically meaningless. For $\rm
\gamma p \rightarrow N^* \rightarrow n\pi^+$ the isotopic
coefficient is equal to $\sqrt{2/3}$, for $\rm \gamma p
\rightarrow N^* \rightarrow p\pi^0$ it is equal to $-\sqrt{1/3}$.
In case of $\rm \Delta$ photoproduction, the respective isotopic
coefficients are $\sqrt{1/3}$ for $n\pi^+$ and $\sqrt{2/3}$ for
$\rm p\pi^0$.


\begin{figure}[t!] 
\centerline{\epsfig{file=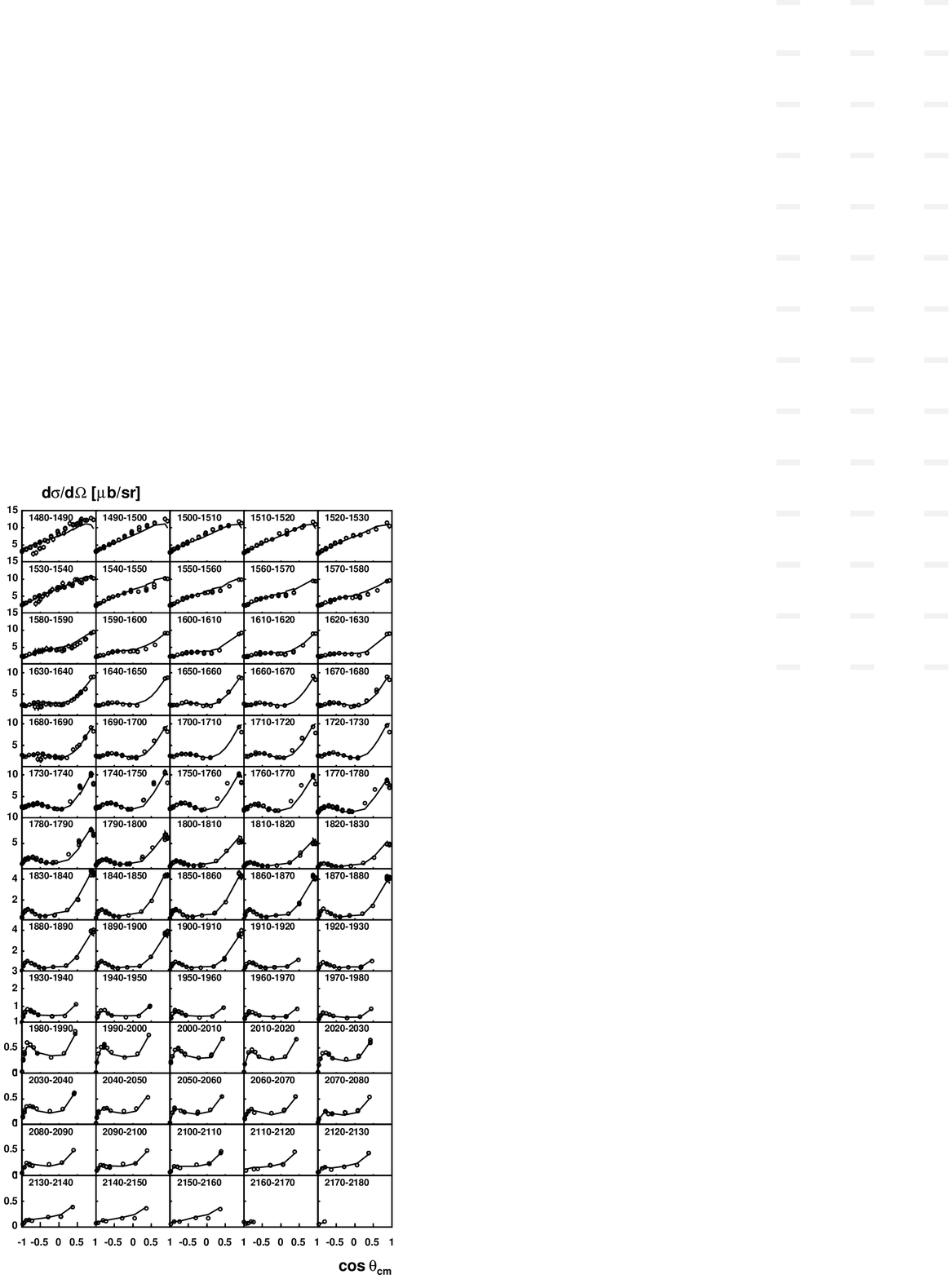,width=0.50\textwidth,clip=}}
\vspace{-1mm}
\caption{Differential cross section for $\rm\gamma p \rightarrow n\pi^+$
from \cite{SAID2} and PWA result (solid line).}
\label{said_piplusn}
\vspace{-4mm}
\end{figure}
Differential cross sections for $\rm \gamma p \rightarrow n\pi^+$
\cite{SAID2} and PWA result are compared in
Fig.~\ref{said_piplusn}. In addition to resonances, a
significant contribution stems from $t$--channel $\pi$ and $\rho$
exchanges (about 10\% and 30\%, respectively). This reaction has a
large number of data points with small statistical errors but the
largest ambiguities in its interpretation. Hence, a small weight
is given to this channel to avoid that it has a significant impact
on baryon masses, widths, or coupling constants. It was only used
to stabilise the fits in case of isospin ambiguities.

\subsection{Fit to the \boldmath $\rm p\eta$ \unboldmath channel}

\begin{figure}[t!] 
\centerline{
\includegraphics[width=0.50\textwidth,height=0.35\textheight,clip]{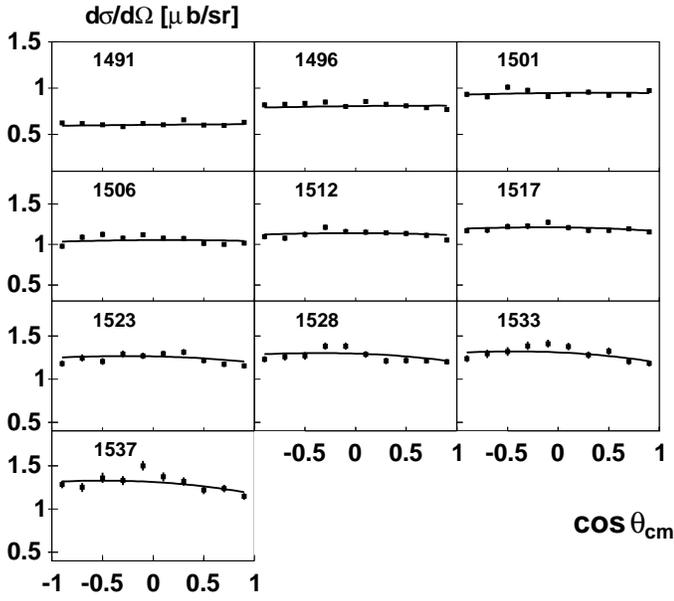}}
\vspace*{-3mm}
\caption{\label{taps98}
Differential cross section for $\rm \gamma p \rightarrow p\eta$ from
Mainz-TAPS data \cite{Krusche:nv} and PWA result (solid line).
}
\vspace*{-1mm}
\end{figure}

Differential cross section for $\rm \gamma p \rightarrow p\eta$ in
the threshold region were measured by the TAPS collaboration at
MAINZ \cite{Krusche:nv}. Data and fit are shown in Fig.~\ref{taps98}. In the
threshold region, the dominant contribution comes from the $\rm
N(1535)$ $\rm S_{11}$ resonance which gives a flat angular distribution.
This resonance strongly overlaps with $\rm N(1650)S_{11}$, and a
two--pole K-matrix parameterisation is used in the fit.

\begin{figure}[pt!] 
\vspace*{-4.5mm}
\centerline{\includegraphics[width=0.50\textwidth,
height=0.92\textheight]{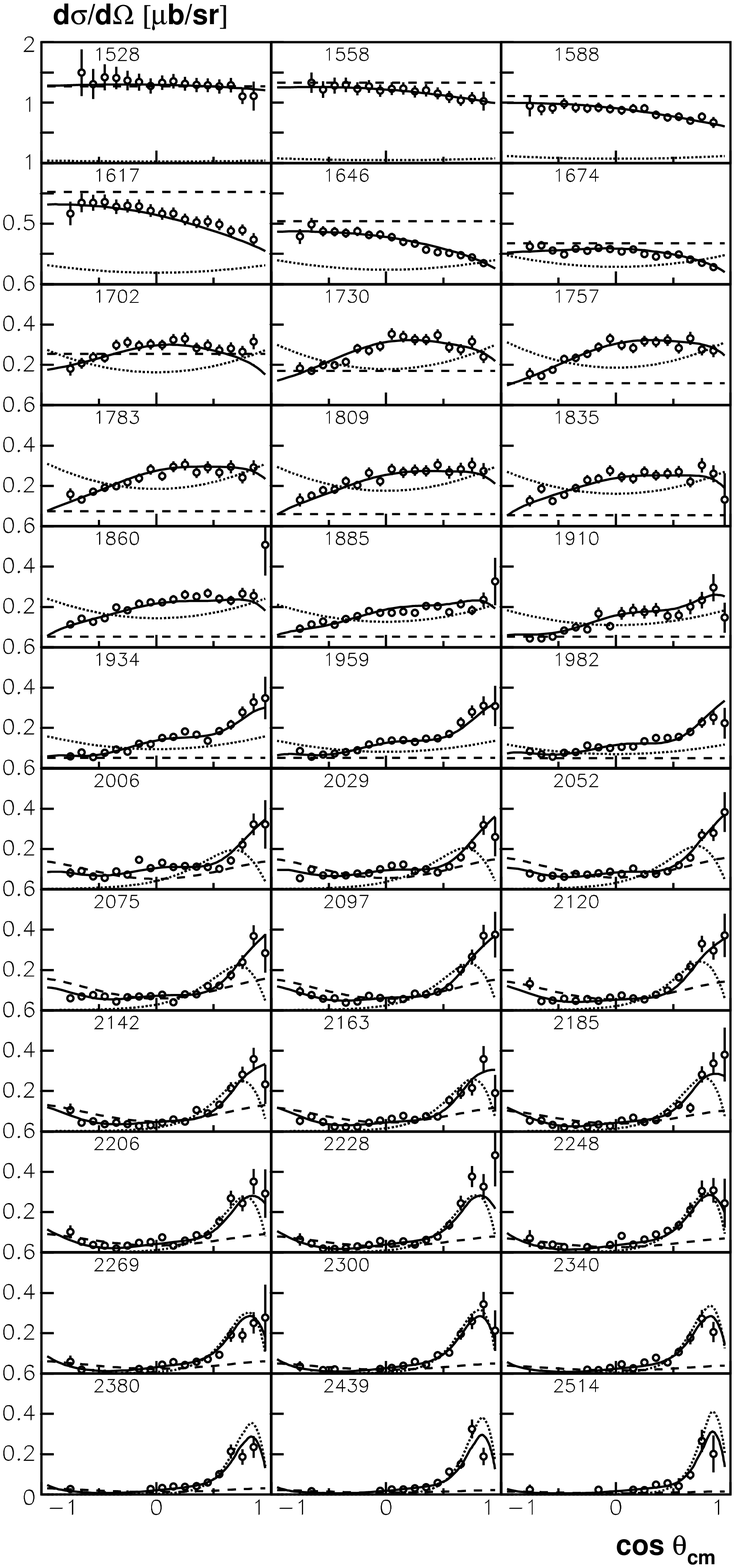}}
\caption{\label{cbelas_eta}
Differential cross section for $\rm \gamma p \rightarrow p\eta$ from
CB-ELSA and PWA result (solid line)~\protect\cite{Crede:04}.
In the mass range below 2\,GeV the
contribution of the two $\rm S_{11}$ resonances is shown as
dashed line and of
$\rm N(1720)P_{13}$ as dotted line. Above 2\,GeV
the contributions of $\rm N(2070)D_{15}$
(dashed line) and $\rho$($\omega$) exchange
(dotted line) are shown.
}
\vspace*{-1cm}
\end{figure}

The CB--ELSA differential cross section~\cite{Crede:04} is given in
Fig.~\ref{cbelas_eta} and compared to the PWA results.  The
contribution of the two $\rm S_{11}$ resonances (dashed line, below
2\,GeV) dominates the $\eta$ production region up to 1650\,MeV. The
most significant further contributions stem from production of
$\rm N(1720)P_{13}$ (dotted line, below 2\,GeV), of $\rm
N(2070)D_{15}$ (dashed line, above 2\,GeV) and $\rho$\,($\omega$)
exchanges (dotted line, above 2\,GeV).

\begin{figure}[t!] 
\vspace*{0.5mm}
\centerline{
\includegraphics[width=0.57\textwidth]{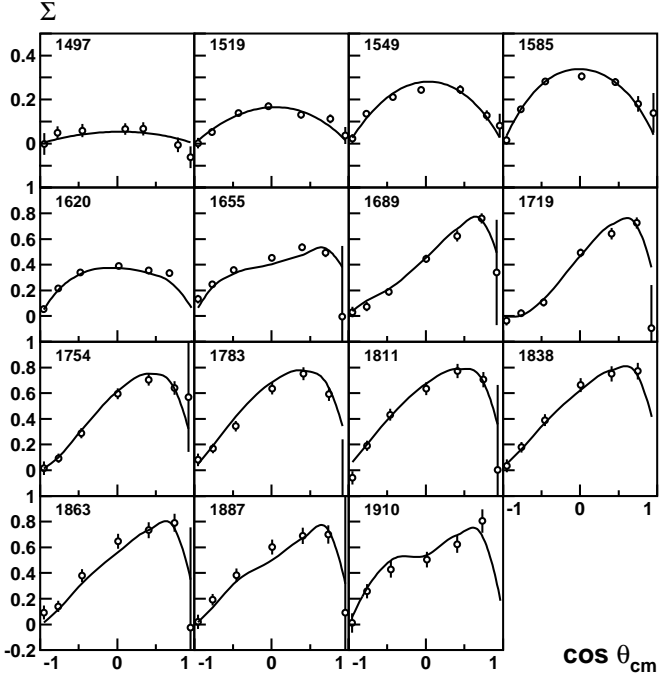}
}
\caption{Photon beam asymmetry $\Sigma$ for $\rm\gamma p \rightarrow
p\eta$ from GRAAL~\cite{GRAAL2} and PWA result (solid line). 
}
\label{graal98}
\vspace*{-8mm}
\end{figure}

Data on the photon beam asymmetry $\Sigma$ for
$\rm \gamma p \rightarrow p\eta$, measured by GRAAL \cite{GRAAL1} are
shown in Fig.~\ref{graal98}.
This data provides essential information on baryon resonances even if
their (p$\gamma$)-- and/or (p$\eta$)--couplings are weak. In addition, the
beam asymmetry data are necessary to determine  the ratio
of helicity amplitudes.


\section{Results}
\subsection{Total cross sections}

From the differential cross sections presented in Figs.~\ref{cbelas_pi0}
and~\ref{cbelas_eta}, absolute cross sections were determined by
integration. The integration is performed by summation of the
differential cross sections  (dots with error bars) and using
extrapolated values for bins with no data, and by integration of the
fit curve.

\begin{figure}[b!] 
\vspace*{-5mm}
\epsfig{file=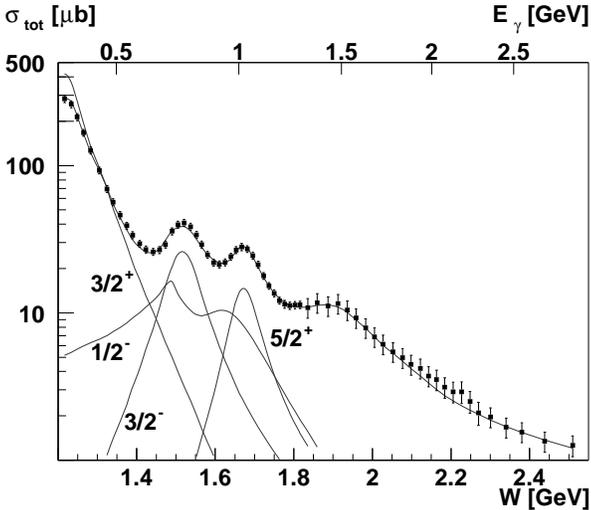,width=0.49\textwidth,clip=}
\vspace*{-5mm}
\caption{\label{figure_pi0tcs} Total cross section (logarithmic
scale) for the reac\-tion $\rm\gamma p\rightarrow p\pi^0$ obtained
by integration of angular distributions of the CB-ELSA data and
extrapolation into forward and backward regions using our PWA
result. The solid line represents the result of the PWA.}
\end{figure}

In the total cross section for $\pi^0$ photoproduction in
Fig.~\ref{figure_pi0tcs}, clear peaks are observed for the first,
second, and third resonance region. With some good will, the
fourth
resonance region can be identified as broad enhancement at
about 1900\,MeV. The decomposition of the peaks into partial waves
and their physical significance will be discussed below.

\begin{figure}[t!]  
\vspace*{-2mm}
\epsfig{file=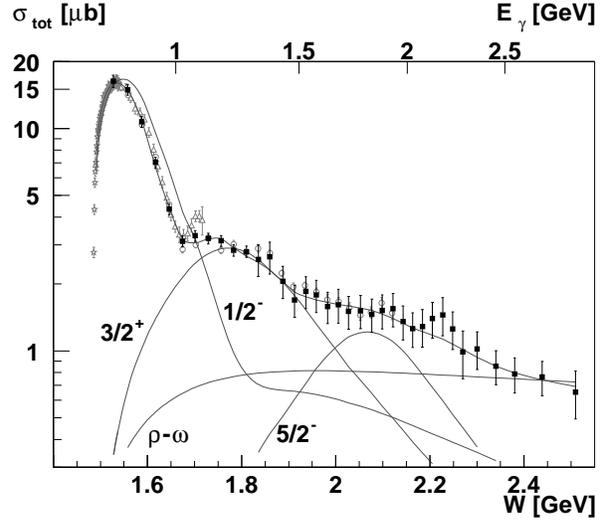,width=0.50\textwidth,clip=}
\vspace*{-5mm}
\caption{Total cross section (logarithmic scale) for the reaction
$\gamma\,\rm{p}\rightarrow\rm{p}\,\eta$
\cite{Crede:04}. Data from other experiments are shown in grey.
The black squares represent the summation over the angular
bins (bins not covered by measurements are taken from the fit), the
solid line represents our fit. The errors are dominantly due to
uncertainties in the normalization. The contributions of the two
S$_{11}$ resonances, of $\rm N(1720)P_{13}$, of $\rm
N(2070)D_{15}$, and of the background amplitudes (mainly
$\rho$($\omega$) exchange) are shown as well.}
\label{total}
\vspace*{-4mm}
\end{figure}

The $\eta$ photoproduction cross section (Fig.~\ref{total})
shows the known strong peak at threshold due to the $S_{11}(1535)$.
The cross section exhibits indications for one further resonance
below 1800\,MeV.

\subsection{The best solution}
The masses and widths of the observed states are presented in
Table~\ref{resonance_list}. Additionally, ratios of helicity
amplitudes $A_{1/2}/A_{3/2}$ and fractional contributions
normalised to the total cross section for the CB--ELSA $\pi^0$-- and
$\eta$-- photoproduction data are included.

A large number of fits (explorative fits plus more than 1000
documented fits) were performed to validate the solution. In these
fits the number of resonances, their spin and parity,
their parameterisation, and the
relative weight of the different data sets were changed.

The errors are estimated from a sequence of fits in which one
variable, e.g. a width of one resonance was changed to series of
fixed values.
All other variables were allowed to adjust freely; the $\chi^2$
changes were monitored as a function of this variable. The errors
given in Table~\ref{resonance_list} correspond to $\chi^2$ changes
of 9, hence to three standard deviations. However, the $3\sigma$
interval corresponds better to the systematic changes observed
when changing the fit hypothesis. 
\par
The resonance properties are compared to PDG values~\cite{PDG}.
Most resonance parameters converge in the fits to values compatible
with previous findings within a 2$\sigma$ intervall of the combined
error. The helicity ratios sometimes seem to be
inconsistent, however they have large errors and the discrepancies are
not really significant.

Three new resonances are necessary to describe the data,
$\rm N(1875)D_{13}$, $\rm N(2070)D_{15}$ and  $\rm N(2200)$ with
uncertain spin and parity. The best fit is achieved for $\rm P_{13}$
quantum numbers. Two further
resonances, $\rm N(1840)P_{11}$ and $\rm N(2170)D_{13}$,
have masses
which are not consistent with established resonances listed by
the PDG. We list them also as new particles. Two resonances,
$\rm N(2000)F_{15}$ and  $\rm\Delta(1940)D_{33}$, are observed for
the first time in photoproduction. PDG mass values for
$\rm N(2000)F_{15}$ range from 1882 to 2175 MeV. We find a mass of
$(1850\pm 25)$\,MeV. Our mass for $\rm\Delta(1940)D_{33}$ is fully compatible
with PDG.
The  $\rm \Delta(1940)D_{33}$ contributes
only at a marginal level. The $\chi^2_{\rm tot}$ changes by
143 units when this resonance is omitted.
The $\rm\Delta(1950)F_{37}$ is observed here at $1893\pm 15$\,MeV
instead of (PDG) $1950\pm{10}$\,MeV.
\par
In this paper we concentrate on the $\rm N(2070)D_{15}$ and
$\rm N(2200)$. The $\rm N(1840)P_{11}$, $\rm N(1875)D_{13}\,$, and
$\rm N(2170)$ $\rm D_{13}$ do not significantly contribute to
$\rm \gamma p\to p\pi^0$, $p\eta$ and
have large couplings to $\rm K\Lambda$ and/or $\rm K\Sigma$.
They will be discussed in \cite{sarantsev}.

Finally a comment is made on resonances with known
photo-couplings but not seen in this analysis. $\rm N(1990)F_{17}$,
\ $\rm \Delta(1600)P_{33}$, \
$\rm \Delta(1910)P_{33}$, $\rm\Delta(1930)D_{35}$,
$\rm \Delta(2420)H_{3\,11}$, and
$\rm N(2190)G_{17}$ are
not observed here. The latter resonance may however be misinterpreted as
$\rm N(2200)P_{13}$ (see Table~\ref{replace}).
The photocouplings of most of these resonances are
seen with weak evidence (one--star rating); only
$\rm\Delta(1600)$ $P_{33}$ has a three--star photo--coupling, and the
$\rm\Delta(1930)D_{35}$ photocoupling
has 2 stars.  We have no explanation why these
states are missing in this analysis. The
$\rm\Delta(1900)S_{31}$, $\rm\Delta(1940)D_{33}$,
and $\rm\Delta(1930)D_{35}$ may form a spin triplet with intrinsic
orbital angular momentum $L=1$ and total spin $S=3/2$ coupling to
$J=1/2,3/2$, and $5/2$ as suggested in \cite{Klempt:2002cu}. Two of these
states are not observed in this analysis. Quark models do not reproduce
these states predicting them to have masses above 2.1\,GeV.
Hence, the question remains open if these states exist at such a low
mass.


\subsection{Significance of resonance contributions}

A systematic study of the significance of new resonances was
carried out. For new resonances the quantum numbers
were changed to any $J^P$ value with $J\leq 9/2$. In the
new fits, all variables were left free for variations
including masses, widths, and couplings of all resonances.
The result of this study is summarised in Table~\ref{pieta_changes}.
The Table illustrates the global deterioration of the fit and the
$\chi^2$ changes for the individual channels. Negative $\chi^2$
changes
indicate that the best quantum numbers are enforced by other
data.
\begin{table*}[pt!]
\caption{Masses, widths and helicity ratio, this analysis.}
\renewcommand{\arraystretch}{1.35}
\begin{center}
\begin{tabular}{lllccccc}
\hline\hline
Resonance& {\footnotesize{ M (MeV)}}   & {\footnotesize{$\Gamma$ (MeV)}}&
 {\footnotesize{$A_{1/2}/A_{3/2}$}} & Fraction & Fraction &
\multicolumn{2}{c}{PDG Rating} \\
&   & &   & $\rm\gamma p \rightarrow p\eta$ & $\rm\gamma p \rightarrow p\pi^0$&overall&N$\gamma$\\
\hline
$\rm N(1440)P_{11}$&$1450\pm 50$&$250\pm 150$    &     &  & 0.007& ~  & \\
PDG &$1440^{+30}_{-10}$&$350\pm 100$   &&&   &****&***\\
\hline
$\rm N(1520)D_{13}$&$1526\pm 4$&$112 \pm 10$&$-0.02\pm 0.10$& 0.030& 0.140 && \\
PDG &$1520^{+10}_{-5}$&$120^{+15}_{-10}$   &$-0.14\pm 0.06$      & ~  &
&****&****\\
\hline
$\rm N(1535)S_{11}$$^*$  & $1530\pm 30$&$210\pm 30$&  & & &&\\
PDG     & $1505\pm 10$&  $170\pm 80$     &   ~ & \multirow{2}*{0.830}&
 \multirow{2}*{0.170} &****&***\\\
$\rm N(1650)S_{11}$$^*$ &$1705\pm 30$&$220\pm 30$&  & &  &&\\
PDG                     &$1660\pm 20$&$160\pm 10 $&  & & &*** &**** \\
\hline
$\rm N(1675)D_{15}$&$1670\pm 20$&$140\pm 40$    &$~~0.40\pm 0.25$& 0.002 & 0.001
&&\\\ PDG & $1675^{+10}_{-5}$ & $150^{+30}_{-10}$& $1.27\pm0.93$& & &****&****\\
\hline
$\rm N(1680)F_{15}$&$1667\pm 6 $&$ 102\pm 15$    &$-0.13\pm 0.05$ & 0.005 &
0.069&&\\ PDG           & $1680^{+10}_{-5}$ & $130\pm 10$ & $-0.11\pm 0.05$ &&& ****&****
\\ \hline
 $\rm N(1700)D_{13}$&$1725\pm 15$&$ 100\pm 15$    &$~~0.45\pm 0.25$& 0.044 &
0.002\\ PDG & $1700\pm 50$ & $100\pm 50$ & $9.00\pm 6.5$& & &***&**\\
\hline $\rm
N(1720)P_{13}$&$1750\pm 40$&$380\pm 40$    &$~~1.5\pm 1.1$& 0.400 &
0.016&   &  \\
PDG & $1720^{+30}_{-70}$& $250\pm 50$ & $-0.9\pm 1.8$ &&&***&** \\
\hline
$\rm N(1840)P_{11}$&$1840^{+15}_{-40}$&$ 140^{+30}_{-15}$    &   & 0.029 & 0.003& new&new
\\ PDG & $1720\pm 30$& $100^{+150}_{\ -50}$ &  &&& ***&*** \\
 \hline
$\rm N(1875)D_{13}$&$1875\pm 25$&$ 80\pm 20$    &$~~1.20\pm 0.45$&
0.013 & 0.000 &new & new\\
\hline
$\rm
N(2000)F_{15}$&$1850\pm 25$&$225\pm 40$    &$~~0.13\pm 1.10$& 0.010 & 0.004&&new\\
PDG           & $\sim 2000$&               &                &       &&**&  \\
 \hline
$\rm N(2070)D_{15}$&$2060\pm 30$&$340\pm 50$    &$~~1.10\pm 0.30$
& 0.195 & 0.012&new&new\\
\hline
$\rm N(2170)D_{13}$&$2166^{+25}_{-50}$&$ 300\pm 65$
&$-1.40\pm 0.80$ & 0.003 & 0.002&new&new\\
PDG           & $\sim 2080$&               &         &     &&**&* \\
\hline
$\rm N(2200)P_{13}$&$2200\pm 30$&$190\pm 50$
&$~~-0.35\pm 0.40$& 0.015  & 0.000&new&new\\
\hline
\hline
 $\rm\Delta(1232)P_{33}$$^{\diamond}$&$1235\pm 4 $&$140\pm
12$    &$~~0.44\pm 0.06$& &0.709 && \\ PDG &$1232\pm 2$&$120\pm5$   &
$~~0.53\pm0.04$  & ~  & &****&****\\
\hline
$\rm\Delta(1620)S_{31}$&$1635\pm 6 $&$106\pm 12$&
& &0.023&& \\ PDG &$1620^{+55}_{-5}$&$150\pm30$   &   & ~  && ****&***\\
\hline
$\rm\Delta(1700)D_{33}$&$1715\pm  20$&$240\pm 35$    &$~~1.15\pm 0.25$& &0.056
&&\\ PDG &$1700^{+70}_{-30}$&$300\pm100$   & $~~1.2^{+0.6}_{-0.4}$  & ~  &&****&*** \\
\hline
$\rm\Delta(1905)F_{35}$&$1870\pm 50$&$370\pm 110$    &$ > 10 $& &0.001
&& \\ PDG &$1905^{+15}_{-35}$&$350^{+90}_{-70}$   & $~~-0.6^{+0.4}_{-0.9}$  & ~
&&****&*** \\ \hline
$\rm\Delta(1920)P_{33}$&$1996\pm 30$&$380\pm 40$    &$~~0.45\pm 0.20$& &0.050  &&\\
PDG &$1920^{+50}_{-20}$&$200^{+100}_{-50}$   & $~~1.7^{+7.}_{-1.0}$  &
~  & &****&*\\
\hline
$\rm\Delta(1940)D_{33}$&$1930\pm 40$
                           &$200\pm 100$    &$~~0.20\pm 0.40$& &0.010 &&new \\
PDG & $\sim 1940$&  &  & &&*& \\
 \hline
$\rm\Delta(1950)F_{37}$&$1893\pm 15 $&$240\pm 30$    &$~~0.75\pm 0.11$&
&0.027  &&\\
PDG &$1950\pm{10}$& $ 300^{+50}_{-10}$   & $~~0.8\pm 0.2$  & ~  & &****&****\\
\hline\hline
\end{tabular}\\
\end{center}
\footnotesize{\hspace*{20mm}$^*$ $K$--matrix fit, pole position of
the scattering amplitude in the complex plane, fraction for the
total \\
\hspace*{22mm} $K$--matrix contribution}\\
\footnotesize{\hspace*{20mm}$^{\diamond}$ This contribution
includes non--resonant background.~~~~~~~}
\renewcommand{\arraystretch}{1.0}
\label{resonance_list}
\end{table*}
\begin{table}[pt!]
\caption{
\label{pieta_changes}
Changes in $\chi^2$ when one of the new resonances is omitted
or replaced by a resonance with different spin and parity $J^P$.
The changes are given for the $\chi^2_{\rm tot}$ (\ref{chi_tot})
and the $\chi^2$ contributions for individual final states calculated
analogously.}
\begin{footnotesize} \renewcommand{\arraystretch}{1.2}
\begin{center}
\begin{tabular}{c|ccccc}
\hline\hline
Resonance
&\multicolumn{5}{c}{$\rm N(2070)D_{15}$}\\
\hline
$J^P$& $\Delta\chi^2_{\rm tot}$ &
\hspace*{-1mm}$\Delta\chi^2_{p\pi^0}$\hspace*{-1mm} &
\hspace*{-1mm}$\Delta\chi^2_{p\eta} $\hspace*{-1mm} &
\hspace*{-1mm}$\Delta\chi^2_{\KL} $\hspace*{-1mm} &
\hspace*{-1mm}$\Delta\chi^2_{\KS} $\hspace*{-1mm} \\
omitted & 1588 & 940 & 199 & 94 & 269 \\
repl. by $1/2^-$ & 1027 & 669 & 128 & 111 &-45  \\
repl. by $3/2^-$ & 1496 & 851 & 214 & -46 & 157  \\
repl. by $7/2^-$ & 1024 & 765 & 108 &  -1 & 19  \\
repl. by $9/2^-$ &  872 & 656 & 112  & -9  & 118  \\
repl. by $1/2^+$ &  832 & 674 & 115  &  55  & 33   \\
repl. by $3/2^+$ & 1050 & 690 & 141 & -42  & 20   \\
repl. by $5/2^+$ &  766 & 627 & 113 & 48  & 123   \\
repl. by $7/2^+$ & 807  & 718 & 112 & -67  & 215  \\
repl. by $9/2^+$ & 1129 & 847 & 131 &  7  & -9 \\
\hline\hline
Resonance
&\multicolumn{5}{c}{$\rm N(2200)P_{13}$}\\
\hline
$J^P$& $\Delta\chi^2_{\rm tot}$ &
\hspace*{-1mm}$\Delta\chi^2_{p\pi^0}$\hspace*{-1mm} &
\hspace*{-1mm}$\Delta\chi^2_{p\eta} $\hspace*{-1mm} &
\hspace*{-1mm}$\Delta\chi^2_{\KL} $\hspace*{-1mm} &
\hspace*{-1mm}$\Delta\chi^2_{\KS} $\hspace*{-1mm} \\
omitted & 190 &  1 &  37 & 43 & 20 \\
repl. by $1/2^-$ & 46 & -18 & 10  & 40 & 0  \\
repl. by $7/2^-$ & 10 & -10 & 7 & 23 & 17   \\
repl. by $9/2^-$ & 18 & -82 &  8  & 16 & 16  \\
repl. by $1/2^+$ & 50 & -8 & 9  & 26 & 42  \\
repl. by $5/2^+$ & 17 & -15  & 10   & 21 & 5  \\
repl. by $7/2^+$ & 13 & -13 & 13  & -10 & 18  \\
repl. by $9/2^+$ & 19 & -9  & 5  & 14 & 17 \\
\hline\hline
\end{tabular}
\end{center}
\renewcommand{\arraystretch}{1.0}
\end{footnotesize}
\label{replace}
\vspace{-.6cm}
\end{table}
\begin{figure}[b!]  
\vspace*{-8mm}
\centerline{\epsfig{file=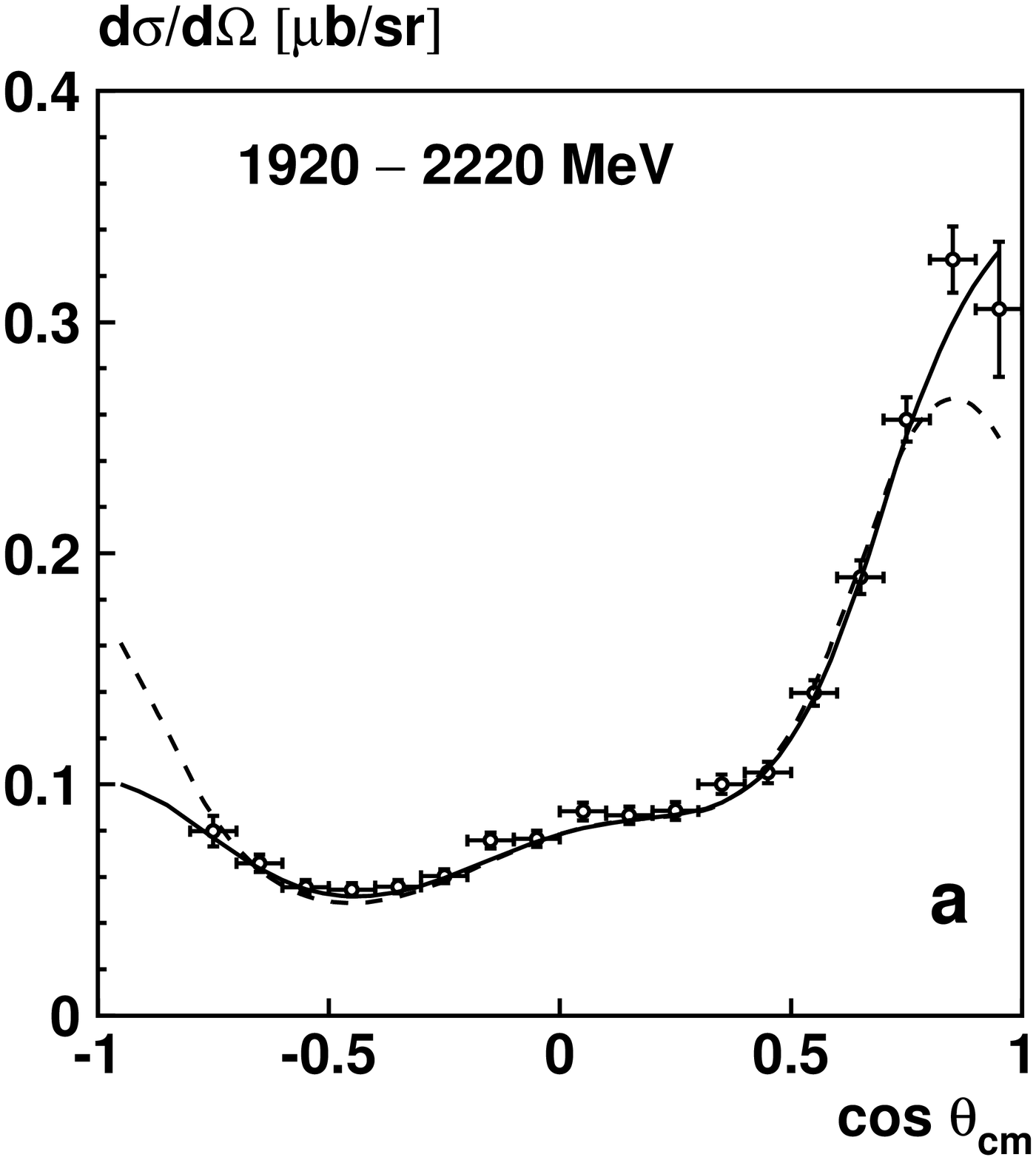,width=0.24\textwidth,clip=}
            \epsfig{file=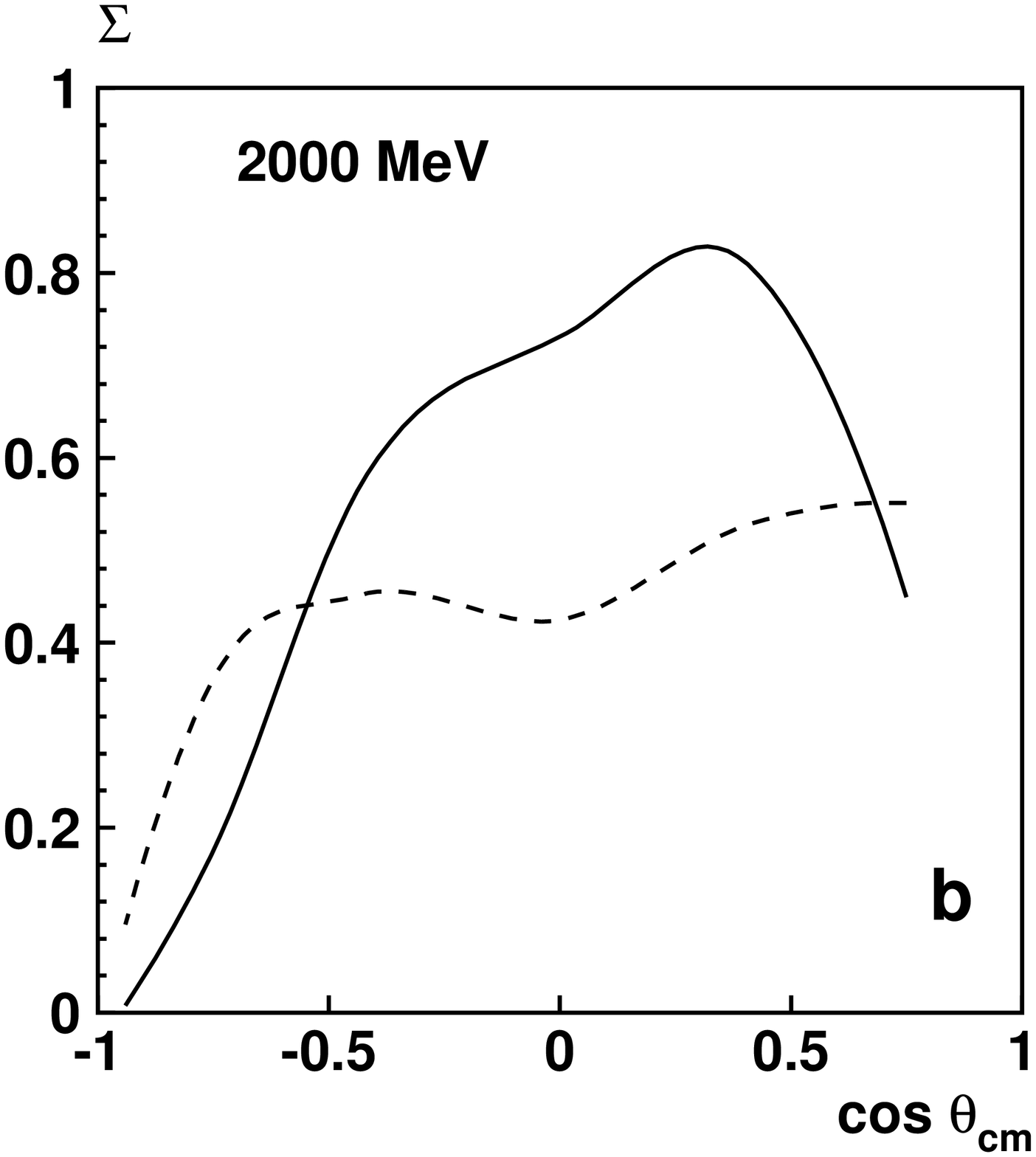, width=0.24\textwidth,clip=}}
\centerline{\epsfig{file=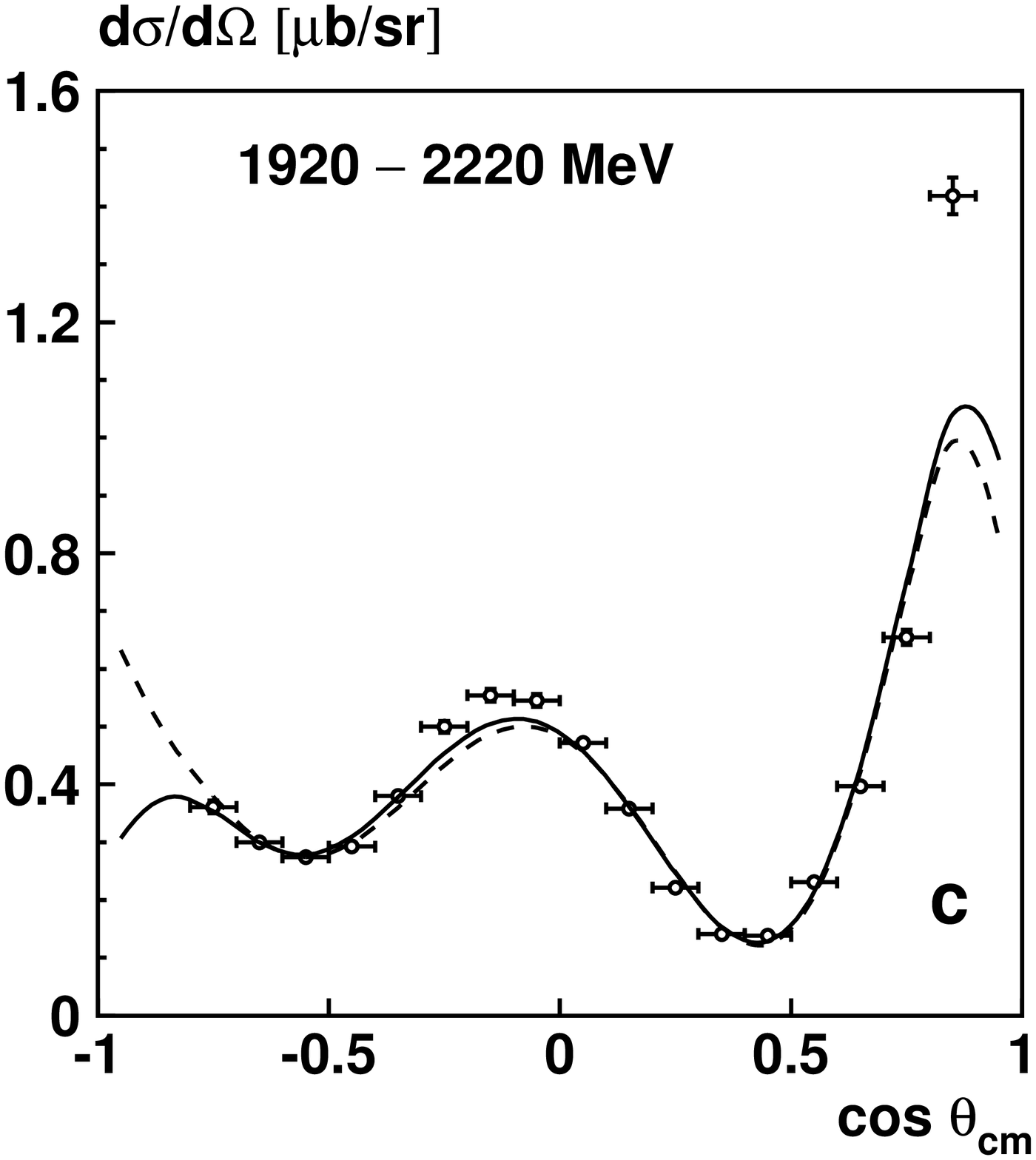,width=0.24\textwidth,clip=}
            \epsfig{file=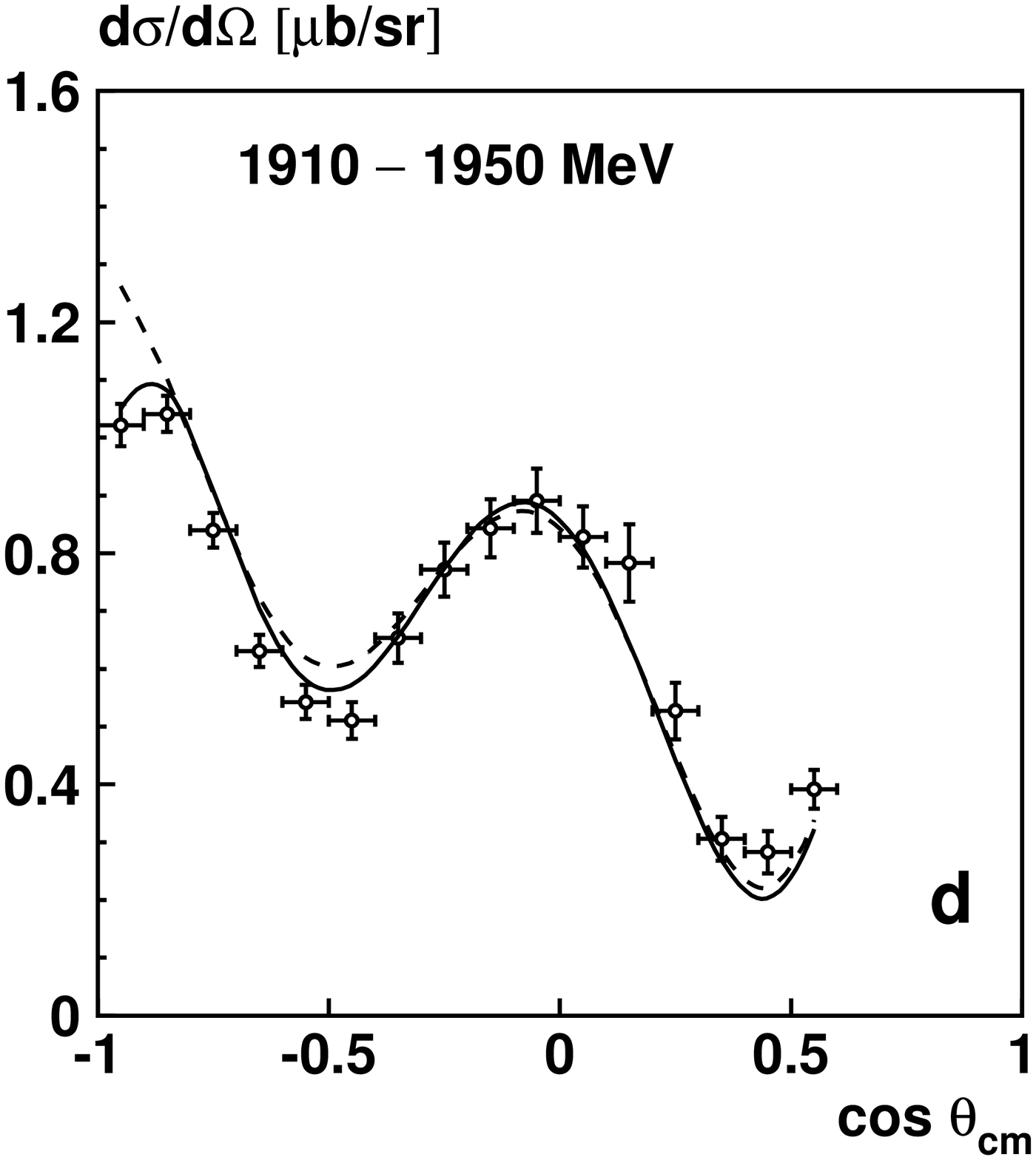,width=0.24\textwidth,clip=}}
\caption{\label{cbelas_chk}
Differential cross section (a),
beam asymmetry (b, predicted curves) from the
reaction $\rm\gamma p \rightarrow p\eta$ and differential cross
sections for $\pi^0$ photoproduction from CB--ELSA (c) and GRAAL05 (d).
Our best PWA fit with $\rm N(2070)D_{15}$ is shown as solid line, 
the dotted line shows a fit when the $5/2^-$ resonance is replaced
by a $7/2^-$ state.  }
\centerline{
\epsfig{file=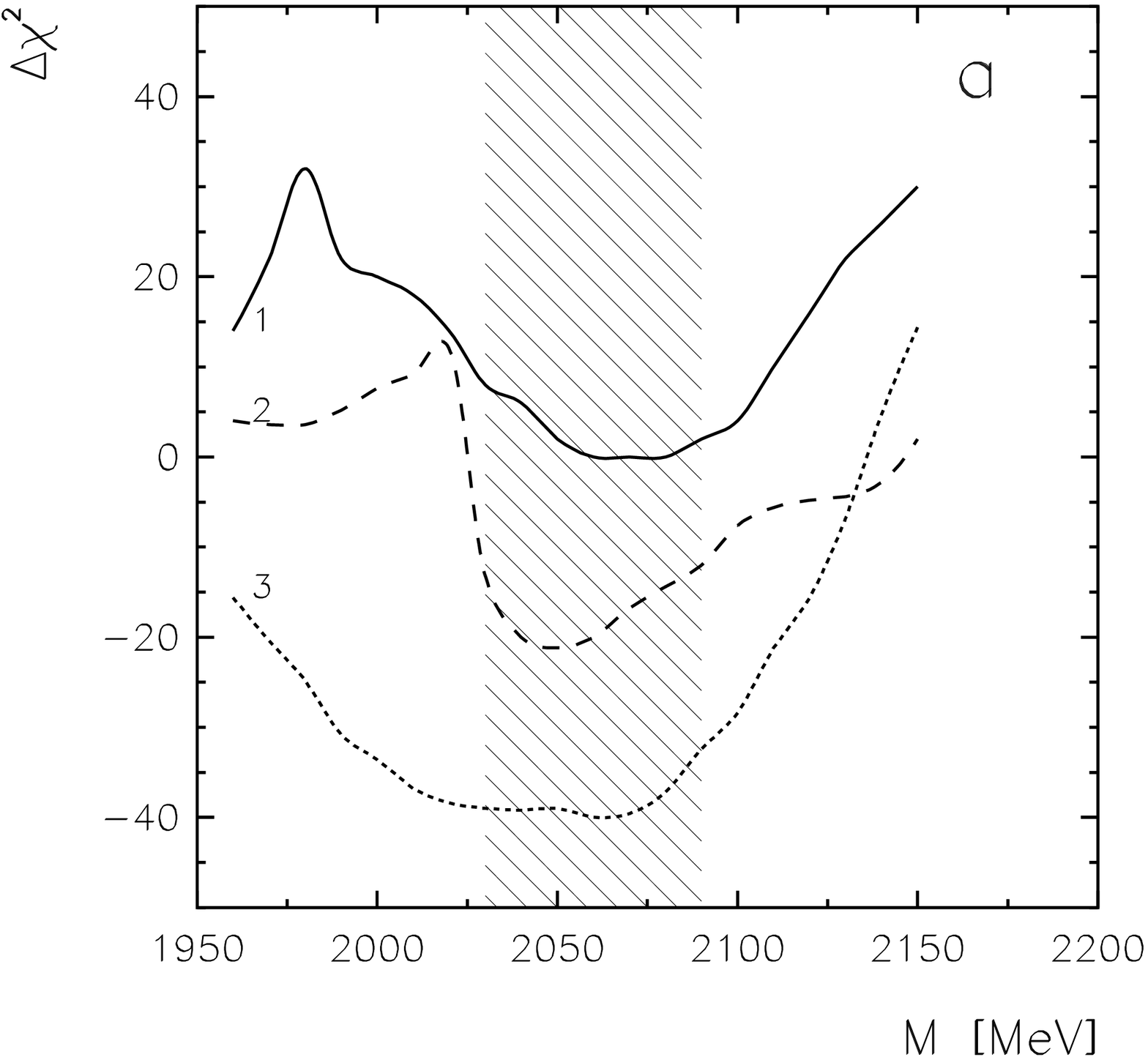,width=0.27\textwidth}\hspace*{-4mm}
\epsfig{file=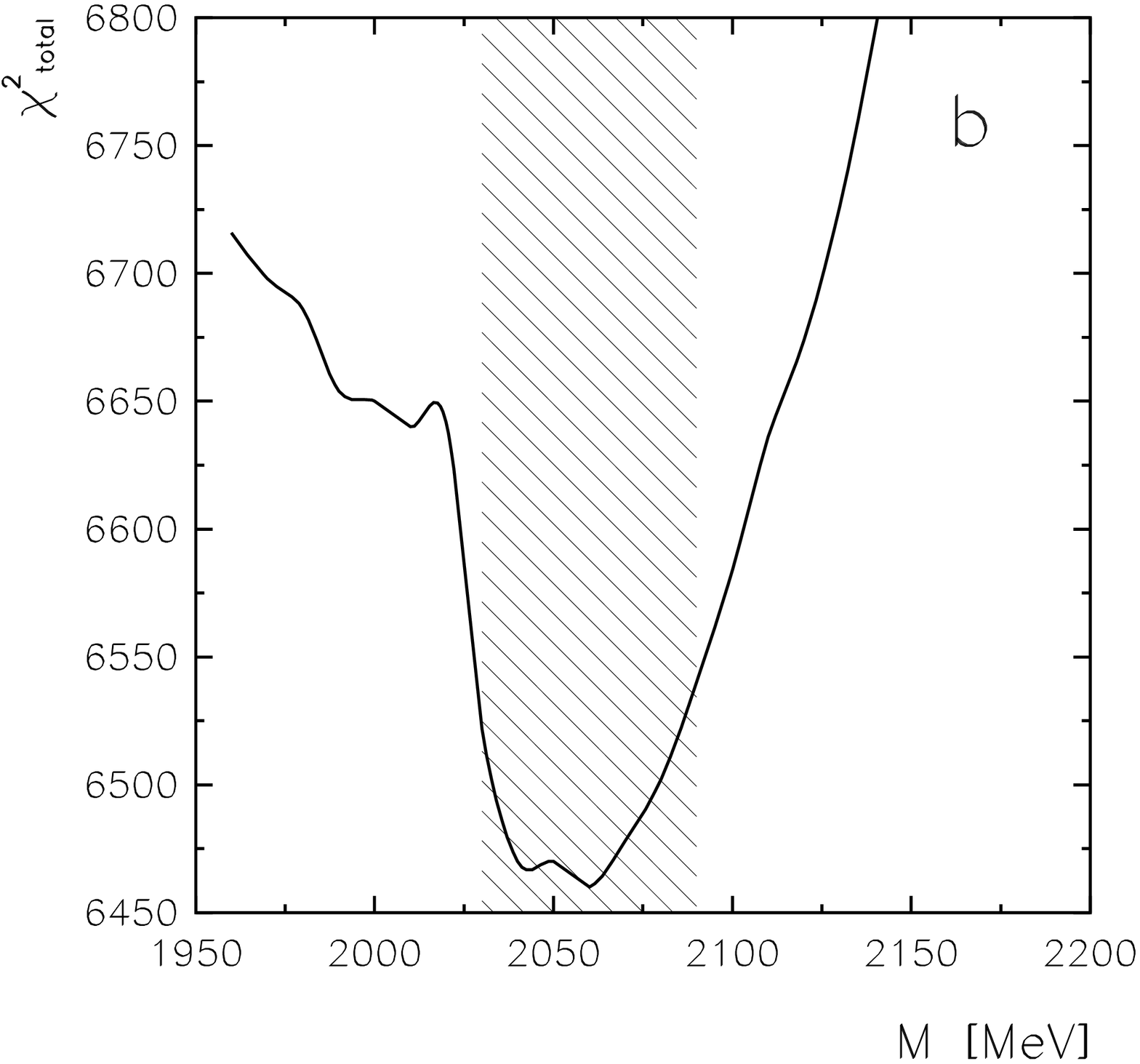,width=0.27\textwidth}\vspace*{-2mm}
}
\caption{The result of $\rm D_{15}(2070)$ mass scan:
a) 1 -- $\rm d\sigma/d\Omega$ for $\rm\gamma p\to p\eta$ (CB-ELSA),
2 -- sum of all reactions with $\KL$ final state
multiplied with 1/5,
3 -- sum of all reactions with $\KS$ final state
multiplied with 1/5,
b) the total $\chi^2$ for all reactions shown in a).}
\label{fig:d15_2070}
\end{figure}
\begin{figure}[t!]
\centerline{
\epsfig{file=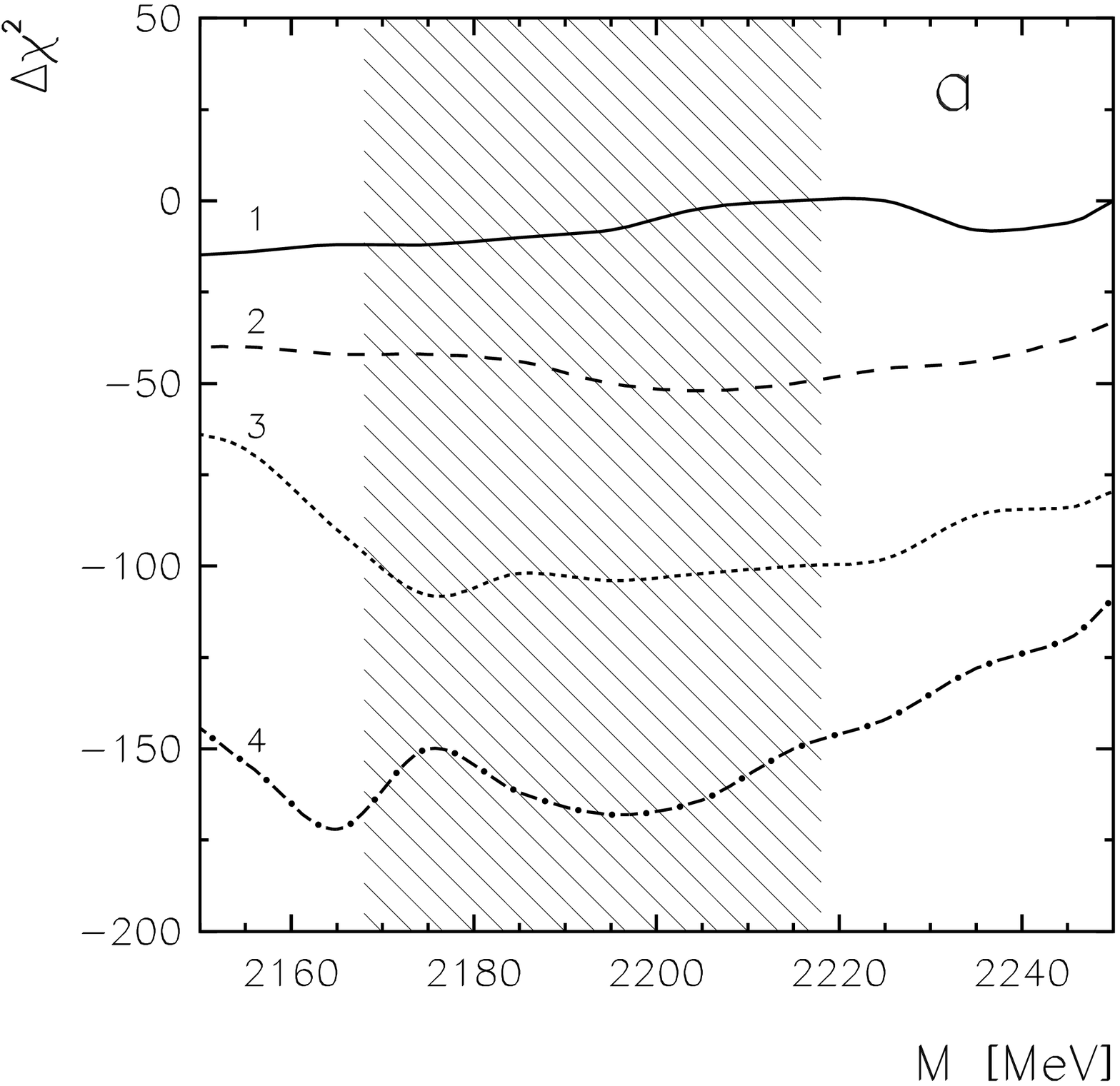,width=0.27\textwidth}\hspace*{-4mm}
\epsfig{file=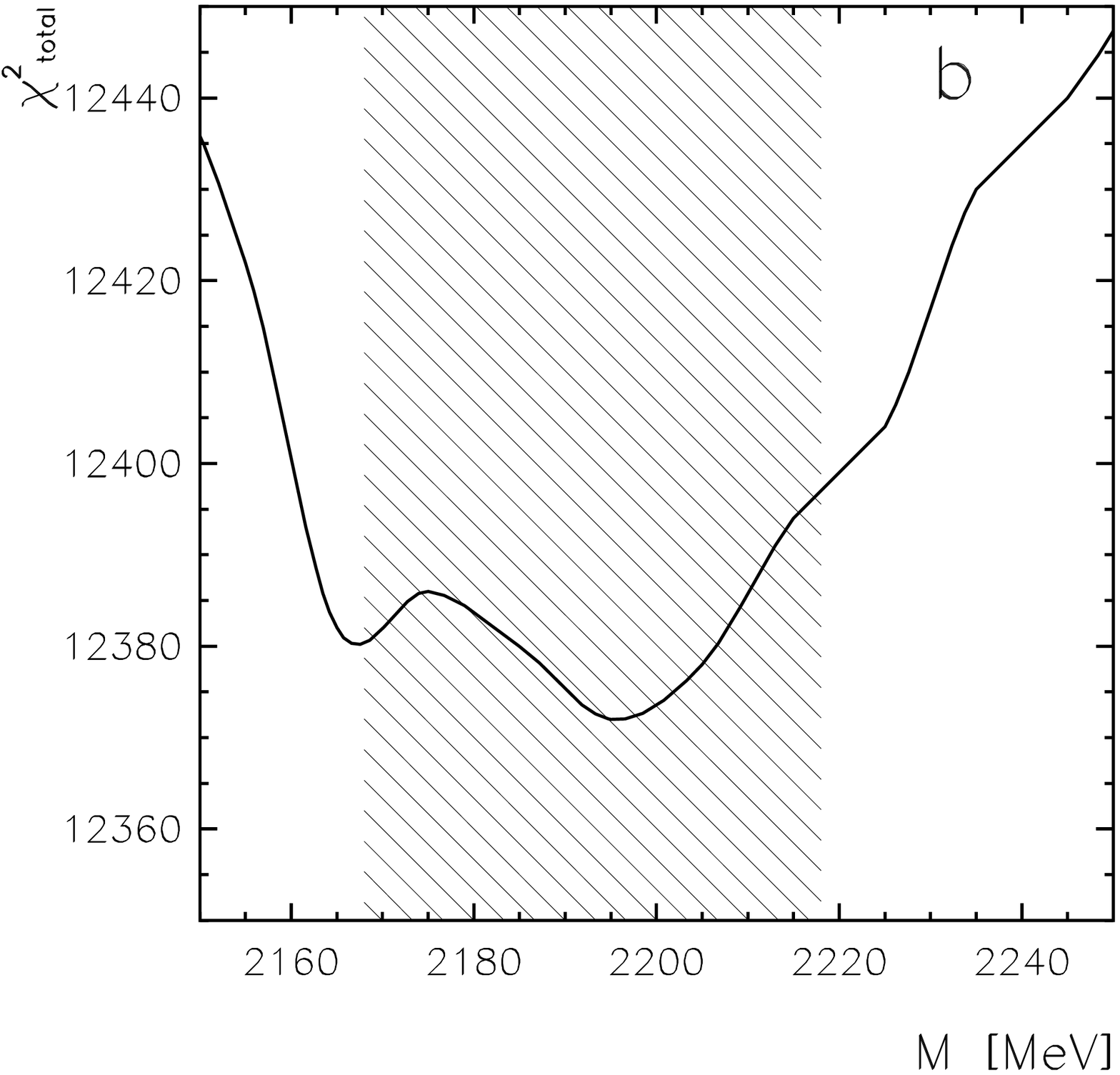,width=0.27\textwidth}\vspace*{-2mm}
}
\caption{The result of $\rm P_{13}(2200)$ mass scan:
a) 1 -- $\rm d\sigma/d\Omega$ for $\rm\gamma p\to p\pi^0 $ (CB-ELSA),
2 -- $\rm d\sigma/d\Omega$ for $\rm\gamma p\to p\eta$ (CB-ELSA),
3 -- sum of all reactions with $\KL$ final state
4 -- sum of all reactions with $\KS$ final state
b) the total $\chi^2$ for all reactions shown in a).}
\label{fig:p13_2200}
\vspace*{-0.5cm}
\end{figure}

The $\rm N(2070)D_{15}$ is the most significant new resonance.
Omitting it changes $\chi^2_{\rm tot}$ by 1589, by 199 for the data on
$\eta$ photoproduction  and by 940 for the data on
$\pi^0$ photoproduction.
Replacing the $J^P$ assignment from $5/2^-$ to $1/2^\pm$,~...,
$9/2^\pm$, the $\chi^2_{\rm tot}$ deteriorates by more than 750.
The deterioration of the fits is visible in the comparison
of data and fit.
One of  the closest description for $\eta$ photoproduction
was obtained fitting with a $7/2^-$
state. In this case, Figs.~\ref{cbelas_chk}\,a,b show
the fits of the differential cross section in the region of resonance
mass and description of the beam asymmetry for highest energy bin.
The shape of the differential cross section at small angles is
close in both cases however the $7/2^-$ state failed to describe
the very forward two points. The beam asymmetry also clearly favours
the $5/2^-$ state.
The $\pi^0$ photoproduction cross sections measured by CB--ELSA
are visually not too sensitive to $5/2^-$ and $7/2^-$ quantum numbers
(see Fig.~\ref{cbelas_chk}\,c) but
there is a clear difference between the two descriptions in the
very backward region. The latest GRAAL results on the $\rm p\pi^0$
differential cross
section which were obtained after discovery of the $\rm N(2070)D_{15}$
\cite{Bartholomy:04}
confirmed $5/2^-$ as favoured quantum numbers
(see Fig.~\ref{cbelas_chk}\,d).

The mass scan of the $\rm D_{15}(2070)$ resonance
($\chi^2$ as a function of the assumed  $\rm D_{15}$ mass)
is shown in Fig.~\ref{fig:d15_2070}.
In the scan, the mass of the  $\rm D_{15}$ was fixed at
a number of values covering the region of interest while
all other fit parameters were allowed to adjust newly.
The sum of $\chi^2$ for $\pi^0$ photoproduction data (CB-ELSA,
GRAAL\,05) does not show any minimum in this region; the destributions
are very flat. Fig.~\ref{fig:d15_2070}a shows separately the sum of $\chi^2$
contributions from the CB--ELSA differential cross section plus the
GRAAL 04
polarisation data, and the sum of the
$\chi^2$ for all $\KL$ and all $\KS$
reactions. A clear minimum is seen in all three data sets.
The sum of $\chi^2$ for all
these reactions is given in Fig.~\ref{fig:d15_2070}b. The shaded area
corresponds to the mass range assigned to this resonance, $(2060\pm
30)$\,MeV.  We conclude that
the  $\rm D_{15}(2070)$ is identified in its decays into $\rm N\eta$,
$\KL$ and $\KS$. Its coupling to  $\rm N\pi$ is weak, hence it is
not surprising that it was not observed in pion induced reactions.

The $\rm N(2200)$  resonance is less significant. Omitting
$\rm N(2200)$ from the analysis, changes  $\chi^2$  for the
CB-ELSA data on $\eta$ photoproduction by 56, and by 20 for
the $\pi^0$--photoproduc\-tion data. Other quantum numbers 
than the preferred $\rm P_{13}$ 
lead to marginally larger $\chi^2$ values.
The mass scan for this state is shown in Fig.~\ref{fig:p13_2200}.
The photoproduction data on $\rm d\sigma/d\Omega$ from CB-ELSA
does not show any minimum, $\eta$ photoproduction data exhibit a
shallow minimum slightly above 2200 MeV.
The sum of all $\KL$ and $\rm K\Sigma$ reactions also have a minimum
in this mass region. The sum of $\chi^2$ for all these reactions is
shown in Fig.~\ref{fig:p13_2200}\,b and from this distribution the
resonance mass can be well defined.

\subsection{The four resonance regions}

The first resonance region dominates pion photoproduction and is due to
the excitation of the $\rm\Delta(1232)P_ {33}$. Its fractional contribution
to $\rm \gamma p\to p\pi^0$ (Table~\ref{resonance_list}) exceeds 1.
There is strong destructive interference between
$\rm\Delta(1232)$ $\rm P_ {33}$, the $\rm P_{33}$ nonresonant amplitude and
$u$--channel exchange.
In the fit without latest GRAAL data on the cross section and beam
asymmetry \cite{GRAAL1}
the $A_{1/2}/A_{3/2}$ helicity ratio of
excitation of the
$\rm\Delta(1232)P_ {33}$ was found to be $0.52\pm 0.06$ which
agrees favorably with the PDG average $0.53\pm 0.04$.
With the new GRAAL05 data included,
this value shifted to $0.44\pm 0.06$.
The $\rm N(1440)P_ {11}$ Roper resonance provides a small
contribution of about 1--3\% compared to the
$\rm \Delta(1232){\rm P}_{33}$.


In the $\rm p\pi^0$ final state
$\rm N(1520)D_{13}$ and the two $\rm S_{11}$ resonances
yield contributions of similar strengths to the second
resonance region. This is consistent with the known photocouplings and
$\rm p\pi$ branching fractions of the three resonances.



The third bump in the $\rm p\pi^0$ total cross section
is due to three major contributions: the
$\rm\Delta(1700)D_{33}$ resonance provides the largest fraction
($\sim\! 35$\%) of the peak, followed by $\rm N(1680)F_{15}$  ($\sim\!
25$\%) and $\rm N(1650)S_{11}$ ($\sim\! 20$\%)
as extracted from the $K$--matrix parameterisation; obser\-ved as
well are the
$\rm\Delta(1620)S_{31}$ ($\sim\! 7$\%)
and $\rm N(1720)P_{13}$  ($\sim\! 6$\%) resonances.
The latter contributes to $\rm p\eta$ with a surprisingly
large fraction; about 90\% of the resonant intensity in this mass
region is assigned to $\rm N(1720)P_{13}\to p\eta$ decays.

In the fourth resonance region we identify
$\rm\Delta(1950)F_{37}$ contributing $\sim\! 41$\%
to the enhancement and $\rm\Delta(1920)P_{33}$ with
$\sim\! 35$\%. Additionally,
the fit requires the presence of
$\rm\Delta(1905)F_{35}$ and $\rm\Delta(1940)D_{33}$.
The high--energy region is dominated by $\rho$($\omega$) exchange in the
$t$ channel as can be seen by the forward peaking in the
differential cross sections.




\subsection{Discussion}

Four new resonances are found in this analysis. The question arises
of course why these resonances have not been found before.
$\rm N(2070)D_{15}$ has a large coupling to $\rm N\eta$ and may
therefore have escaped discovery. The $\rm N(1875)D_{13}$
and $\rm N(2170)D_{13}$ states couple strongly to the $\rm K\Lambda$
and $\rm K\Sigma$
channels; the existence of the first state has already
been suggested in~\cite{Mart:1999ed} from an analysis of older
SAPHIR data on  $\rm \gamma p\to K\Lambda$~\cite{Tran:1998qw}.
Cutkosky~\cite{Cutkosky:1980rh} reported
two $\rm N\,D_{13}$  resonances at $(1880\pm 100)$ and $(2081\pm
80)$\,MeV with respective widths of $(180\pm 60)$ and $(300\pm
100)$\,MeV. The $\rm N(1840)P_{11}$ appears in all channels.
The evidence for it is discussed in \cite{sarantsev}.
The $\rm N(2200)$ does not have such characteristic features. 
It improves the description of the data in a difficult
mass range and further data will be required to establish or to
disprove its existence. Its preferred quantum numbers are $\rm P_{13}$
but it seems not unlikely that $\rm N(2200)$ should be identified
with  $\rm N(2190)G_{17}$ (which gives the second best PWA solution).  


The three largest contributions to the $\eta$ photoproduction cross
section stem from $\rm N(1535)S_{11}$, $\rm N(1720)P_{13}$, and
\begin{figure}[b!]
\includegraphics[width=0.48\textwidth]{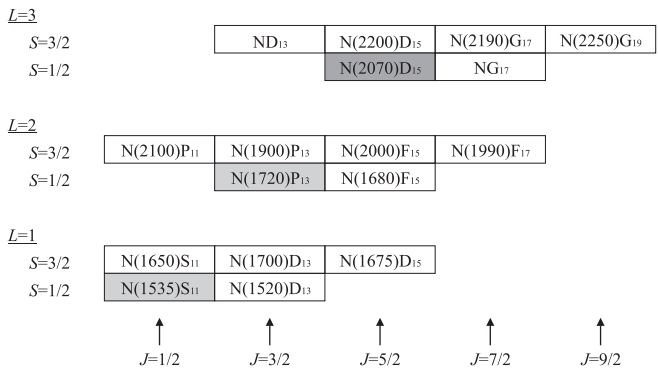}
\caption{\label{ls}
N$^*$ resonances with quantum numbers which can be assigned to
orbital angular momentum excitations with $L=1,2,3$. The
quark spin, $S=1/2$ or $S=3/2$, and the orbital angular momentum
couple to the total spin $J$. Note that mixing between states of the
same parity and total angular momentum is possible. Resonances with
strong coupling to the N$\eta$ channel are marked in grey.}
\end{figure}
$\rm N(2070)D_{15}$. We tentatively assign
($J\!=\!1/2; L\!=\!1,S\!=\!1/2$)
quantum numbers to the first state; $\rm N(1720)P_{13}$
and $\rm N(1680)F_{15}$ form a spin doublet, hence the dominant
quantum numbers of $\rm N(1720)P_{13}$ must be
($J\!=\!3/2; L\!=\!2, S\!=\!1/2$). Thus
it is tempting to assign ($J\!=\!5/2; L\!=\!3, S\!=\!1/2$)
quantum numbers to
$\rm N(2070)D_{15}$. The three baryon resonances with strong
contributions to the $\rm p\eta$ channel thus all have spin $S=1/2$ and
orbital and spin angular momenta adding antiparallelly with $J\!=\!L-1/2$.
Fig.~\ref{ls} depicts this scenario.
\par
The large $\rm N(1535)S_{11}\to N\eta$ coupling has been a topic of a
controversial discussion. In the quark model, this coupling arises
naturally from a mixing of the two ($J\!=\!1/2;L\!=\!1,S\!=\!1/2$) and
($J\!=\!1/2;L\!=\!1,S\!=\!3/2$) harmonic-oscillator
states~\cite{Isgur:1978xj}.
However, $\rm N(1535)S_{11}$ is very close to the $\rm K\Lambda$ and
$\rm K\Sigma$ thresholds and the resonance can be understood as
originating from coupled--channel meson--baryon chiral dynamics~
\cite{Kaiser:1995cy}. Alternatively, the strong  $\rm N(1535)S_{11}\to
N\eta$ coupling  can be explained as delicate interplay
between confining and fine structure
interactions \cite{Glozman:1995tb}.
\par
A consistent picture of the large $\rm N(1535)S_{11}\to N\eta$
coupling should explain the systematics of $\rm N\eta$ couplings. We
note a kinematical similarity: The three resonances with large
$\rm N\eta$ partial decay widths are those for which
the dominant intrinsic orbital excitation $L=1,2,3$ and the 
decay orbital angular momenta $\ell=0,1,2$ are related
by $J=L-1/2=\ell+1/2$. The intrinsic quark spin configuration remains in
a spin doublet. 
\section{Summary}
We have presented a partial wave analysis of data on photoproduction
of  $\rm \pi N$, $\rm \eta N$,
$\rm K\Lambda$, and $\rm K\Sigma$ final states. The data include total
cross sections 
and angular distributions, beam asymmetry measurements as
well as the recoil polarisation in case of hyperon production.
A reasonable description of all data was achieved by introducing
14 $\rm N^*$ and seven $\rm\Delta^*$ resonances.
\par
Most baryon resonances are found with masses, widths and ratios of helicity
amplitudes which are fully compatible with previous findings.
New resonances are required to fit the data, $\rm N(1840)P_{11}$,\,
$\rm N(1875)D_{13}$,\, $\rm N(2070)D_{15}$,\, $\rm N(2170)D_{13}$,\,
and\, $\rm N(2200)$. The\, $\rm N(1840)P_{11}$\, resonance\,
could, however, be identical with $\rm N(1710)P_{11}$ and
$\rm N(2170)$ $\rm D_{13}$ with $\rm N(2080)D_{13}$.
\par
Three resonances are found to have very large couplings to
$\rm N\eta$, $\rm N(1535)S_{11}$,
$\rm N(1720)P_{13}$, and $\rm N(2070)D_{15}$. The dynamical
origin of this preference remains to be investigated.

\subsection*{Acknowledgements}
We would like to thank the CB-ELSA/TAPS Collaboration
for numerous discussions on topics related to this work.
We acknowledge
financial support from the Deu\-tsche Forschungsgemeinschaft
within the SFB/TR16.
The collaboration with St.Petersburg received funds from the
DFG and the Russian Foundation for Basic Research.
U.~Thoma thanks for an Emmy Noether grant from the DFG.
A.V. Anisovich and A.V.~Sarantsev
acknowledge support from the Alexander von Humboldt Foundation.

\end{document}